\documentclass[useAMS,usenatbib]{mn2e}

\usepackage{times}
\usepackage{color}
\usepackage{graphicx}
\usepackage{mathptmx}
\providecommand{\LyX}{L\kern-.1667em\lower.25em\hbox{Y}\kern-.125emX\@}
\usepackage{natbib}

\def\farcs{\hbox{$.\!\!^{\prime\prime}$}}

\title[Where do long-period comets come from?]{Where do long-period comets come from?
\newline 26 comets from the non-gravitational Oort spike}
\author[ M. Kr\'olikowska and P.A. Dybczy\'{n}ski]{Ma\l gorzata Kr\'olikowska$^1$\thanks{E-mail:
mkr@cbk.waw.pl} and Piotr A. Dybczy\'{n}ski$^{2}$\thanks{E-mail: dybol@amu.edu.pl}\\
$^1$Space Research Centre of the Polish Academy of Sciences,
Bartycka 18A, 00-716 Warsaw, Poland; \\
$^2$Astronomical Observatory Institute,
A.Mickiewicz Univ., Sloneczna 36, 60-286 Pozna\'{n}, Poland,}
\begin{document}

\date{}

\pagerange{\pageref{firstpage}--\pageref{lastpage}} \pubyear{2002}

\maketitle

\label{firstpage}

\begin{abstract}
Since 1950, when Oort published his hypothesis several important new
facts were established in the field. In this situation the apparent source region (or
regions) of long-period comets as well as the definition of the
dynamically new comet are still open questions, as well as the
characteristics of the hypothetical Oort Cloud. The aim of this investigation is to look for the
apparent source of selected long period comets and to refine the definition of
dynamically new comets.
\newline Basing on pure gravitational original orbits all comets studied
in this paper were widely called dynamically new. We show, however,
that incorporation of the non-gravitational forces into the orbit
determination process significantly changes the situation.
\newline We determined precise non-gravitational orbits of all investigated comets
and next followed numerically their past and future motion during
one orbital period. Applying ingenious Sitarski's
(\citeyear{sitarski:1998}) method of creating swarms of virtual
comets compatible with observations, we were able to derive the
uncertainties of original and future orbital elements, as well as
the uncertainties of the previous and next perihelion distances.
\newline We concluded that the past and future evolution of cometary orbits under
the Galactic tide perturbations is the only way to find which comets are
really dynamically new. In our sample less than 30\% of comets are in fact
dynamically new. Most of them had small previous perihelion distance.
On the other hand, 60\% of them will be lost on hyperbolic orbits in the
future. This evidence suggests that the apparent source of long-period
comets investigation is a demanding question.
\newline We also have shown that a significant percentage of long-period comets can
visit the zone of visibility during at least two or three consecutive
perihelion passages.
\end{abstract}


\section{Introduction}\label{sec:Introduction}

Jan Oort\citeyearpar{oort:1950} proposed a hypothesis, that the
Solar System is surrounded by a huge, spherical cloud of comets, now
called the Oort Cloud. As the argument for the existence of such a
cloud he completed a list of 19 original orbits\footnote{In this
paper we define original and future orbits as barycentric osculating
orbital elements derived at a distance of 250\,AU from the Sun.} of
well observed long-period (hereafter LP) comets and showed, that
their inverse of semi-major axis $1/a$ has the distribution
apparently peeked near zero, at the positive side. He concluded that
new long-period comets come from distances of 50\,000 to 150\,000
AU. He also showed, that perturbations by passing stars can change
cometary orbit significantly, making it observable as
{}``dynamically new'' long-period comet. During the last 60 years
several new factors occurred in this field.

First, the population of precise original cometary orbits increased
to several hundreds. In the recently published, 17th edition of the
Catalogue of Cometary Orbits \citep{marsden-cat:2008} (hereafter
MWC08) 499 precise original and future cometary orbits are listed.
Second, new and dominating perturbing force must be included in the
model: tidal action of the Galactic disk and Galactic centre. They
are negligible when calculating original and future orbits but when
following the past or future motion of a comet at larger distances
one must take them into account. It means, that looking at the
original orbit elements we cannot tell the past or predict the
future motion of this body, for example foretell the previous/next
perihelion or aphelion distances. This was demonstrated by
\citeauthor{dyb-hist:2001} (\citeyear{dyb-hist:2001,dyb-hab3:2006}).
Third, we learned how to investigate and in many cases we can
successfully determine non-gravitational forces in the motion of
long-period comets. This can significantly change our knowledge of
original and future orbits (see for example:
\citealp{marsden-sek-ye:1973};
\citealp{krolikowska:2001,krolikowska:2006a}). It also means, that
tens of precise original orbits with determined non-gravitational
forces could be added to the MWC08 list.

Taking all this into account, the apparent source region (or
regions) of long-period comets as well as the definition of the
dynamically new comet are still open questions, as well as the
characteristics (if not the existence) of the hypothetical Oort
Cloud. In this paper we start to investigate these problems
concentrating on 26 comets constituting so-called non-gravitational
(hereafter NG) Oort spike, i.e. having $1/{\rm a}_{{\rm
ori,NG}}<10^{-4}$\,AU$^{-1}$. This particular choice comes from the
fact, that after applying non-gravitational model in  orbit
determination we obtain significantly different distribution of the
inverse original semi-major axes. Almost all $1/a_{\rm ori}$ values were shifted
to larger values and among them a lot of hyperbolic cases changed
into elliptic ones. Since we want to study the implications of
obtaining non-gravitational original orbits on the Oort hypothesis,
the past (and future) motion of these 26 comets were analyzed at
first.

In Section \ref{sec:Observations-and-their} we describe
observational material and its processing. In the next section
calculation of original and future cometary orbits for the selected
26 non-gravitational Oort spike comets is described in detail.
Section \ref{sec:Past-and-future} describes the numerical
integration of the past and future motion of these comets under the
action of Galactic tides. Then, results and conclusions are
presented.

\section{Observations and their processing}\label{sec:Observations-and-their}

MWC08 includes 184 comets with $1/{\rm a}_{{\rm
ori}}<10^{-4}$\,AU$^{-1}$ where for majority of them the orbits are
determined with the highest accuracy. At first we looked at 154 most
precise orbits. Thus, we took into account all comets with the
quality class 1 according to the classification introduced by
\citet{mar-sek-eve:1978}. About half of them have perihelion
distances less than 3.0~AU, what means that we should suspect
detectable influence of the non-gravitational forces. The inverse
dependence of the strength of non-gravitational forces on the
perihelion distance is widely expected and indeed is clearly visible
in our Fig.\ref{fig:spik_ng} (to be described in detail in the next
section). We took into account almost all these objects (except ten
comets discovered before 1950 where positional data were not
available for us) and additionally all LP~comets with the NG~orbits
in MWC08 (29 objects, $1/{\rm a}_{{\rm ori}}$ are not given for them
in the Catalogue). In majority of cases, the observational material
taken for orbital determinations is more complete than in the MWC08:
for comets from the XIX century the observations were collected
directly from the published papers, for comets from the first half
of XX century the data were collected in Warsaw in cooperation with
the Slovakian group at the Astronomical Institute in Bratislava and
Tatranska Lomnica, more modern data were taken from the Minor Planet
Electronic Circulars available through the Web Page at
http://www.cfa.harvard.edu.

\begin{table}
\caption{\label{tab:Obs-mat}Observational material for 26 comets under consideration}
\begin{tabular}{lcc}
\hline
 &  & No. \tabularnewline
Name  & Observational arc  & obs. \tabularnewline
\hline
C/1885 X1 Fabry           & 1885 12 02 -- 1886 07 20  & 228 \tabularnewline
C/1892 Q1 Brooks          & 1892 09 01 -- 1893 07 13  & 191 \tabularnewline
C/1913 Y1 Delavan         & 1913 10 25 -- 1915 09 07  & 1009 \tabularnewline
C/1940 R2 Cunningham      & 1940 08 25 -- 1941 04 01  & 370 \tabularnewline
C/1946 U1 Bester          & 1946 11 01 -- 1948 10 02  & 142 \tabularnewline
C/1952 W1 Mrkos           & 1952 12 10 -- 1953 07 18  & 36 \tabularnewline
C/1956 R1 Arend-Roland    & 1956 11 08 -- 1958 04 11  & 249 \tabularnewline
C/1959 Y1 Burnham         & 1960 01 04 -- 1960 06 17  & 88 \tabularnewline
C/1978 H1 Meier           & 1978 04 28 -- 1979 12 09  & 287 \tabularnewline
C/1986 P1 Wilson          & 1986 08 05 -- 1989 04 11  & 688 \tabularnewline
C/1989 Q1 Okazaki-Levy-R. & 1989 08 24 -- 1989 12 24  & 231 \tabularnewline
C/1989 X1 Austin          & 1989 12 06 -- 1990 06 27  & 281 \tabularnewline
C/1990 K1 Levy            & 1990 05 21 -- 1992 04 01  & 678 \tabularnewline
C/1991 F2 Helin-Lawrence  & 1991 02 23 -- 1992 09 30  & 114 \tabularnewline
C/1993 A1 Mueller         & 1992 11 26 -- 1994 08 17  & 746 \tabularnewline
C/1993 Q1 Mueller         & 1993 08 16 -- 1994 04 17  & 526 \tabularnewline
C/1996 E1 NEAT            & 1996 03 15 -- 1996 10 12  & 249 \tabularnewline
C/1997 BA$_{6}$ Spacewatch& 1997 01 11 -- 2004 09 15  & 529 \tabularnewline
C/1997 J2 Meunier-Dupouy  & 1997 05 05 -- 1999 10 09  & 1446 \tabularnewline
C/1999 Y1 LINEAR          & 1999 10 29 -- 2003 07 19  & 884 \tabularnewline
C/2000 SV$_{74}$ LINEAR   & 2000 09 05 -- 2005 05 12  & 2189 \tabularnewline
C/2001 Q4 NEAT            & 2001 08 24 -- 2006 08 18  & 2661 \tabularnewline
C/2002 E2 Snyder-Murakami & 2002 03 08 -- 2003 01 08  & 941 \tabularnewline
C/2002 T7 LINEAR          & 2002 10 12 -- 2006 03 07  & 4451 \tabularnewline
C/2003 K4 LINEAR          & 2003 05 28 -- 2006 11 17  & 3658 \tabularnewline
C/2004 B1 LINEAR          & 2004 01 28 -- 2008 08 24  & 2057 \tabularnewline
\hline
\end{tabular}
\end{table}

>From these class~1 objects we were able to derive the sample of
26~LP~comets with determinable non-gravitational effects from the
astrometric data and with the non-gravitational reciprocals of
semi-major axes not greater than $10^{-4}$\,AU$^{-1}$ ($1/a_{{\rm
ori,NG}}<10^{-4}$\,AU$^{-1}$, hereafter: NG~Oort spike). Among them
only 9~comets (C/1885~X1, C/1956~R1, C/1959~Y1, C/1986~P1A,
C/1989~Q1, C/1990~K1, C/1993~A1, C/2001~Q4, C/2002~T7) have NG
orbits presented in MWC08. C/1990~K1 Levy has formally $1/a_{{\rm
ori,NG}}$ slightly greater than the adopted limit but the
uncertainty interval of it evidently overlap with the NG~Oort spike
defined above. We have not included in our sample objects which are
still potentially observable because the derived orbits for these
objects could not be treated as definitive results. The preliminary
investigation of the original non-gravitational reciprocals of
semi-major axes, $1/a_{{\rm ori,NG}}$, are given in
\citet{krolikowska:2006a} for the majority of these 26~comets. In
that paper data have not been weighted, while now, to obtain the
more reliable orbits, we decided to include the weighting of
observations into the data processing.

The determination of the NG~parameters in the motion of LP~comets is
very difficult mainly due to limited observational material covering
one apparition or even just half apparition (in the case when comet
was discovered after its perihelion passage). Thus, the processing
of astrometric data is crucial for this purpose.
\citet{krolikowska:2006a} divided each set of astrometric data into
pre-perihelion and post-perihelion arc (when it was possible) and
then selected data in each part independently. The difference of rms
for the whole data set (for the pure gravitational solution) and the
rms obtained separately for both parts of the orbit indicates how
strongly the NG-effects affect the orbital motion. After many tests
we have recognized, however, that the weighting is crucial for the
orbit fitting for comets discovered before 1950 and sometimes for
modern comets, too. Thus, we decided to adopt more advanced data
treatment. Each set of astrometric data has been processed
individually for the pure gravitational orbit and non-gravitational
case independently in the following way.

%
\begin{figure*}
\begin{centering}
\includegraphics[width=8.0cm]{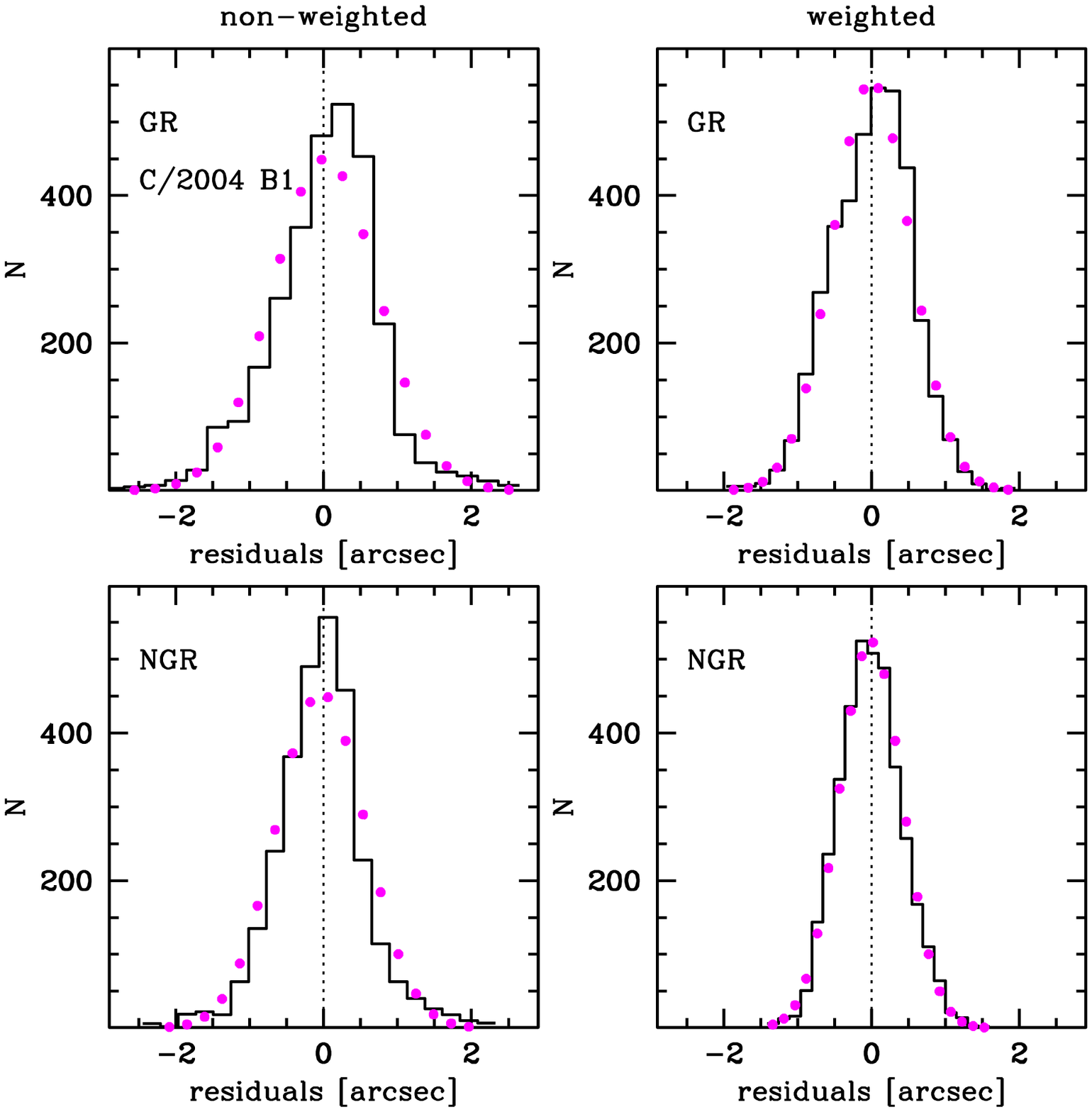} \includegraphics[width=8.0cm]{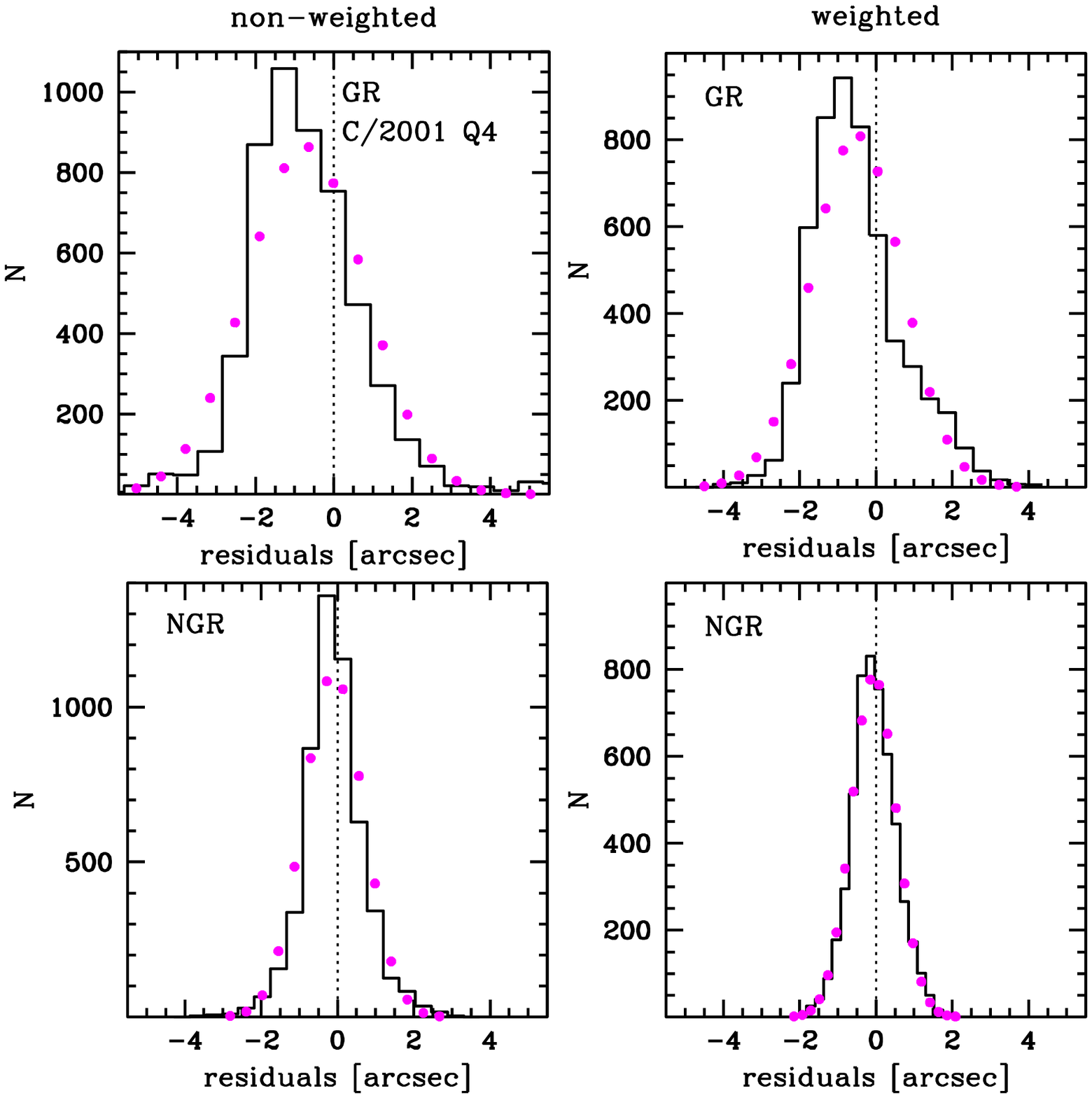}
\par\end{centering}

\caption{The O-C distributions for Comet C/2004~B1~LINEAR and
C/2001~Q4~NEAT. The non-weighted data are presented in the left-side
column for each comet and the weighted data -- in the right column
where the gravitational cases are displayed in upper figures and the
NG~cases in lower figures. The best-fitting Gaussian distributions
are shown by dots, each chosen exactly in the middle of histogram
bin.}

\begin{centering}
\label{fig:04b1gau}
\par\end{centering}
\end{figure*}

For each comet we tested two methods of selecting and weighting data
for both: pure gravitational and non-gravitational orbit,
independently. It means that we determined four nominal solutions
for each object: two for pure gravitational case and two for
non-gravitational case (based on different data treatment). It
allows us to choose the best and as homogeneous as possible data
processing for the whole sample of LP~comets. We tried to use the
most but reasonable discriminative selection criterion and according
to our tests the year 1950 can serve as a suitable dividing date.
Thus, we divided all considered comets into two sets: comets
discovered before and after 1950. These two sets of observations
were refined independently for gravitational and non-gravitational
orbit determination by applying

\vspace{0.2cm}

\noindent \textit{for comets before 1950:}

\begin{itemize}
\item Chauvenet's criterion%
\footnote{short description of both criterions is given in \ref{app:Cha}\label{fn:repeat1}%
} for selection procedure $+$ weighting,
\item Bessel's criterion\footnotemark[\value{footnote}] 
 for selection procedure $+$ weighting,
\end{itemize}

\vspace{0.2cm}

\noindent  \textit{for comets after 1950:}

\begin{itemize}
\item independent Bessel's selection for observations before and
after perihelion, no weighting,
\item Bessel's criterion for selection procedure $+$ weighting,
\end{itemize}
\noindent To reduce systematic errors in the observational material,
such as the bias associated with a site as a function of time, one
should consider specific procedures. For example, in the case of
long sequences of observations we divided the whole observational
material into a few time subintervals according to the internal
structure of material, i.e. according to the existing gaps in
observations.

The data of two comets with  the number of observations less than
100 (C/1952~W1 and C/1959~Y1) have not been weighted. For two of the
comets discovered before 1950 the Chauvenet's and Bessel's
criterions gave the same set of residuals and for the next three
comets the differences in the number of rejected residuals are very
small (less than 1\%). Thus, we present here results derived for
more discriminative Bessel's criterion also for comets before 1950.
One should note that our data processing resulted in reasonably low
number of rejections (below 17\% -- for comets before 1960 and with
mean percentage of rejected data of 3.4\%  -- for comets after 1960;
the only extreme case here is C/2004~B1, for which 9\% of
observations was rejected). The description of observational data
used for orbital determination is presented in
Table~\ref{tab:Obs-mat} where comets are ordered by discovery date.

To determine the NG~cometary orbit we used the standard formalism
proposed by \citet{marsden-sek-ye:1973} where the three orbital components
of the NG~acceleration acting on a comet are symmetric relative to
perihelion:

\begin{eqnarray}
F_{i}=A_{{\rm i}}\cdot & g(r) & ,\qquad A_{{\rm i}}={\rm ~const~~for}\quad{\rm i}=1,2,3,\label{f_r}\\
 & g(r) & =\alpha\left(r/r_{0}\right)^{-m}\left[1+\left(r/r_{0}\right)^{n}\right]^{-k},\label{g_r}\end{eqnarray}

\noindent where $F_{1},\, F_{2},\, F_{3}$ represent the radial,
transverse and normal components of the NG~acceleration,
respectively. The exponential coefficients $m,\, n,\, k$ are equal
to $2.15$, $5.093$, and $4.6142$, respectively; the normalization
constant $\alpha=0.1113$ gives $g(1$~AU$)=1$; the scale distance
$r_{0}=2.808$~AU. From orbital calculations, the NG~parameters
$A_{1},A_{2}$, and $A_{3}$ were derived together with six orbital
elements within a given time interval (numerical details are given
in \citealp{krolikowska:2006a}). This NG~model assumes that water
sublimates from the whole surface of an isothermal cometary nucleus.
One can see in the 10$^{{\rm th}}$ column of
Table~\ref{tab:a_original} that in some cases just two NG~parameters
were determinable with reasonable accuracy (in two cases -- only one
NG~parameter). We decided to use in this investigation the minimal
necessary number of NG~parameters. For example, in the case of
C/1993~A1 and C/1996~E1 the NG~parameter $A_{3}$ is determinable,
however, the decrease of rms (relative to NG~solution with $A_{1}$
and $A_{2}$) is insignificant with no perceptible changes in O-C
diagrams and O-C distributions (compare with online material in
\citet{krolikowska:2004}). The asymmetric NG~model which introduces
the additional NG~parameter $\tau$ -- the time displacement of the
maximum of the $g\left( r(t-\tau) \right) $ relative to perihelion,
seems be important just for one comet, C/1959~Y1~Burnham, what makes
the number of NG~parameters for this comet to be equal 4.

\vspace{0.2cm}




%
\begin{figure}
\begin{centering}
\includegraphics[width=8.8cm]{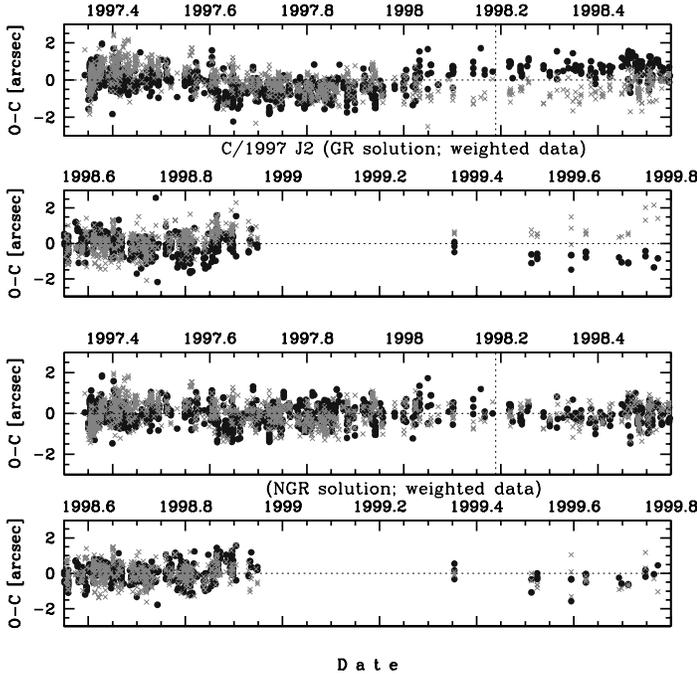}
\par\end{centering}
\caption{The O-C diagrams for Comet C/1997~J2 Meunier-Dupouy
(weighted data). Two upper figures are given for O-C based on pure
gravitational solution, two lower -- for O-C based on NG~solution.
Residuals in right ascension are shown as black dots and in
declination -- as grey crosses; the moment of perihelion passage is
shown by dashed vertical line.}
\begin{centering}
\label{fig:1997j2_OC}
\par\end{centering}
\end{figure}


%
\begin{figure}
\begin{centering}
\includegraphics[width=8.8cm]{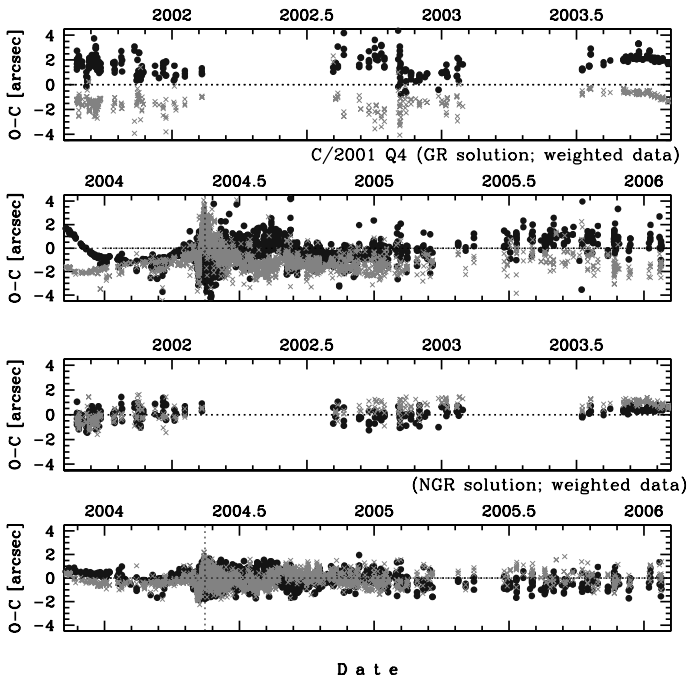}
\par\end{centering}

\caption{The O-C diagrams for Comet C/2001~Q4 NEAT (weighted data). Two upper
figures are given for O-C based on pure gravitational solution, two
lower -- for O-C based on NG~solution (for the clarity the three
last observations taken in August 2006 are not displayed). Symbols
have the same meaning as in Fig.~\ref{fig:1997j2_OC}}

\begin{centering}
\label{fig:2001q4_OC}
\par\end{centering}
\end{figure}


%

To decide whether the derived NG~solution better represents the actual
motion of individual comet than the pure gravitational orbit we use
three criterions:

\begin{itemize}
\item decrease of the rms for the NG~solution in comparison with the pure
gravitational solution,
\item smaller trends in O-C diagrams for NG orbit fitting,
\item the O-C distribution of NG orbit closer to the normal one.
\end{itemize}
After inspection of the O-C diagrams and distributions we always
conclude that the NG~orbit derived from the weighted and selected
data sets represents the actual comet motion much better than the
pure gravitational orbit. This fact is visualized in
Fig.~\ref{fig:04b1gau} for two comets C/2004~B1 and C/2001~Q4.
Therefore, the pure gravitational orbits are presented in this paper
only for comparison and to show that such orbits could produce
misleading orbital results and incorrect conclusions about future
and past history of cometary dynamics. The examples of differences
in the O-C diagrams between pure gravitational and NG~solutions are
presented for two comets in
Figs.~\ref{fig:1997j2_OC}~and~\ref{fig:2001q4_OC}. For comet
C/1997~J2 only pure gravitational orbit is presented in MWC08 as
well as at the JPL Small-Body Database Browser
(http://ssd.jpl.nasa.gov/sbdb.cgi), whereas for Comet C/2001~Q4
NG~solutions are presented in both sources.
Fig~\ref{fig:1997j2_OC} reveals that some trends visible in pure
gravitational O-C diagram for C/1997~J2 Meunier-Dupouy are almost
canceled for NG~solution while the rms
decreased from 0\farcs 91 to 0\farcs 67. Similar behavior was
noticed for all NG~solutions given in this paper even for four
objects with largest perihelion distance (q$>3.0$\,AU) as was shown
in Fig.~\ref{fig:1997j2_OC} for Comet C/1997~J2.

\noindent 
Among investigated sample of NG~Oort spike comets C/2001~Q4~NEAT
presents the case of one of the most significant decrease of rms
from 1\farcs 69 (pure gravitational orbit, non-weighted data) to
0\farcs 63 (NG~orbit, weighted data) (Table~\ref{tab:a_original},
Figs.~\ref{fig:04b1gau}, \ref{fig:2001q4_OC}). In this group are
also C/1959~Y1~Burnham, C/1990~K1~Levy, C/1993~A1~Mueller and
C/2002~T7~LINEAR.

For many comets with evidently detectable NG-effects we obtained
strongly non-Gaussian O-C distribution for pure gravitational orbit
whereas the O-C distribution follows the normal distribution when
the NG~orbit were fitted. An excellent example of such behavior is
Comet C/2004~B1. The departures from the best Gaussian fitting to
O-C residuals are shown in Fig.~\ref{fig:04b1gau}. One can note that
the deviations of O-C distribution from normal distribution in the
pure gravitational case always decreased for weighted data in
comparison to non-weighted data, however in some cases this
deviation is still clearly visible even for weighted data as one can
see for both comets presented in Fig.~\ref{fig:04b1gau}. It seems
evident that non-Gaussian distribution of O-C residuals results from
ignoring of NG~forces in the cometary motion. In a few cases
(C/1990~K1, C/1993~A1, C/2001~Q4, C/2002~T7 and C/2003~K4), however,
the O-C distributions for NG~solution obtained on the basis on
weighted data differ from the Gaussian distribution also. In all
these five cases the residuals in NG~model still display some trends
too (especially around perihelion passage), though less significant
than in the pure gravitational case (Fig.~\ref{fig:2001q4_OC}). We
tested that the time-shifted g(r)-function relative to perihelion
passage do not improved the O-C diagrams here. It seems that some
other function than standard g(r) would be more adequate to describe
the NG~effects in these comets. It should be noticed, however, that
the presented NG~nominal orbits for these five comets are closer to
actual cometary orbit than pure gravitational solutions.

And the last, important remark concerning the gravitational
solutions determined from weighted data. In practice, these
solutions are not purely gravitational since the time distribution
of weights could compensate the NG~trends. For three objects from
our sample (C/2002~T7, C/2001~Q4 and C/1997 BA$_{6}$) relative
decreasing of rms and some time-dependence in weights are evidently
visible . For the most spectacular case (C/2002~T7) the rms of
gravitational solutions decreased from 1\farcs 21 (non-weighted
data) to 0\farcs 61 (weighted data) whereas the rms of NG~solutions
decreased from 0\farcs 64 to 0\farcs 58, respectively, and
systematic trends in weights disappeared. Thus, we should keep in
mind that so called gravitational solutions determined on the basis
of weighted data could partially incorporate NG~trends.

\begin{center}
\begin{table*}
\caption{\label{tab:a_original}Original and future semi-major axes derived
from pure gravitational nominal solutions (columns~2--3) and NG~nominal
solutions (columns~4--5) where the number of NG~parameters determined
for NG~solutions is given in the column~10. The rms's and number
of residuals are given in the columns 6--7 and 8--9, respectively,
where 'GR' refers to gravitational nominal solution.
If there are two rows for a particular comet, the first describes
results with weighting and the second without weighting. Single row
(older comets) means that weighting was applied.}

{\setlength{\tabcolsep}{4.2pt}
\begin{tabular}{@{}lr@{$\pm$}rr@{$\pm$}rr@{$\pm$}rr@{$\pm$}rccrrr@{}}
\hline
{\tiny Name }  & \multicolumn{4}{c}{{ gravitational solutions}} & \multicolumn{4}{c}{{ NG~solutions}} & { rms$_{{\rm GR}}$ }  & { rms$_{{\rm NG}}$ }  & \multicolumn{3}{c}{{ number of}}\tabularnewline
 & \multicolumn{2}{c}{{ $1/{\rm a}_{{\rm ori}}$}} & \multicolumn{2}{c}{{ $1/{\rm a}_{{\rm fut}}$}} & \multicolumn{2}{c}{{ $1/{\rm a}_{{\rm ori}}$}} & \multicolumn{2}{c}{{ $1/{\rm a}_{{\rm fut}}$}} &  &  & { res. }  & { res. }  & { NG}\tabularnewline
 & \multicolumn{4}{c}{{ i n ~~~u n i t s ~~~of ~~~$10^{-6}$\,AU$^{-1}$}} & \multicolumn{4}{c}{{ i n ~~~u n i t s ~~~of ~~~$10^{-6}$\,AU$^{-1}$}} & { \farcs }  & { \farcs }  & { GR }  & { NG }  & { par. }\tabularnewline
\hline
{ 1 }  & \multicolumn{2}{c}{{ 2 }} & \multicolumn{2}{c}{{ 3 }} & \multicolumn{2}{c}{{ 4 }} & \multicolumn{2}{c}{{ 5 }} & { 6 }  & { 7 }  & { 8 }  & { 9 }  & { 10 }\tabularnewline
\hline
{ C/1885 X1 Fabry }          & { ~-2.73}  & { 11.28}  & { ~-256.73}  & { 11.28}  & { +61.04}  & { 17.17} & { -103.42}  & { 41.33 }& { 3.92 }  & { 3.58 }  & { 394}   & { 390}   & { 2}\tabularnewline
{ C/1892 Q1 Brooks }         & { -57.34}  & { 12.52}  & { ~-568.73}  & { 12.52}  & { +58.07}  & { 50.88} & { -486.86}  & { 23.28 }& { 3.20 }  & { 2.80 }  & { 343}   & { 334}   & { 2}\tabularnewline
{ C/1913 Y1 Delavan }        & { +22.14}  & { ~1.30}  & { ~~+56.41}  & { ~1.30}  & { +52.57}  & { ~4.23} & { ~+86.84}  & { 4.23}  & { 2.06 }  & { 2.00 }  & { 1874}  & { 1860}  & { 1}\tabularnewline
{ C/1940 R2 Cunningham }     & { -60.00}  & { 17.66}  & { -1630.62}  & { 17.67}  & { +51.66}  & { 17.30} & { -1494.22} & { 36.52} & { 1.78 }  & { 1.49 }  & { 678}   & { 670}   & { 2}\tabularnewline
{ C/1946 U1 Bester }         & { ~+5.72}  & { ~4.97}  & { ~~+32.53}  & { ~4.97}  & { +25.41}  & { ~5.25} & { ~+52.21}  & { 5.25}  & { 1.53 }  & { 1.35 }  & { 257}   & { 253}   & { 1}\tabularnewline
{ C/1952 W1 Mrkos }          & { -140.69} & { 21.69}  & { ~-298.66}  & { 21.69}  & { ~-0.07}  & { 85.78} & { ~-41.39}  & { 106.94}& { 1.14 }  & { 1.08 }  & { 61}    & { 61}    & { 2}\tabularnewline
{ C/1956 R1 Arend-Roland }   & { -97.97}  & { ~5.46}  & { ~-603.51}  & { ~5.47}  & { +10.37}  & { 11.42} & { -579.69}  & { 9.97}  & { 1.68 }  & { 1.39 }  & { 461}   & { 458}   & { 3}\tabularnewline
{ {} }                       & { -104.38} & { ~5.02}  & { ~-609.92}  & { ~5.03}  & { ~+7.83}  & { 11.19} & { -587.32}  & { 10.59} & { 1.68 }  & { 1.41 }  & { 449}   & { 449}   & { 3}\tabularnewline
{ C/1959 Y1 Burnham }        & { -138.32} & { 15.93}  & { ~-589.97}  & { 15.94}  & { ~+2.09}  & { 123.88}& { -286.65}  & { 33.46} & { 3.00 }  & { 1.60 }  & { 146}   & { 146}   & { 4}\tabularnewline
{ C/1978 H1 Meier  }         & { +23.79}  & { ~2.42}  & { -1028.68}  & { ~2.42}  & { +80.55}  & { 10.73} & { -1025.19} & { 8.15}  & { 1.04 }  & { 1.00 }  & { 565}   & { 565}   & { 2}\tabularnewline
{ {}}                        & { +24.52}  & { ~2.73}  & { -1027.96}  & { ~2.73}  & { +89.23}  & { 12.83} & { -1018.74} & { 8.09}  & { 1.22 }  & { 1.18 }  & { 549}   & { 549}   & { 2}\tabularnewline
{ C/1986 P1 Wilson  }        & { ~+9.53}  & { ~0.82}  & { ~+735.14}  & { ~0.82}  & { +42.00}  & { 1.97}  & { +767.22}  & { 2.35}  & { 1.35 }  & { 1.11 }  & { 1364}  & { 1361}  & { 2}\tabularnewline
{ {}}                        & { ~+0.44}  & { ~0.99}  & { ~+725.17}  & { ~0.99}  & { +42.33}  & { 2.39}  & { +767.22}  & { 2.86}  & { 1.62 }  & { 1.31 }  & { 1353}  & { 1353}  & { 2}\tabularnewline
{ C/1989 Q1 Okazaki-Levy-R.} & { -12.14}  & { 14.02}  & { ~+182.62}  & { 14.02}  & { +42.90}  & { 22.24} & { ~+80.49}  & { 27.32} & { 1.84 }  & { 1.30 }  & { 458}   & { 452}   & { 2}\tabularnewline
{ {}}                        & { ~-6.47}  & { 12.66}  & { ~+188.30}  & { 12.66}  & { +54.58}  & { 24.90} & { ~+93.84}  & { 33.52} & { 1.86 }  & { 1.46 }  & { 441}   & { 438}   & { 2}\tabularnewline
{ C/1989 X1 Austin}          & { +22.39}  & { ~2.07}  & { ~-365.62}  & { ~2.07}  & { +40.84}  & { 9.89}  & { -368.96}  & { 5.42}  & { 1.31 }  & { 1.28 }  & { 537}   & { 537}   & { 2}\tabularnewline
{ {}}                        & { +22.40}  & { ~2.37}  & { ~-365.62}  & { ~2.37}  & { +41.71}  & { 12.53} & { -373.15}  & { 6.77}  & { 1.49 }  & { 1.49 }  & { 530}   & { 530}   & { 2}\tabularnewline
{ C/1990 K1 Levy}            & { +17.59}  & { ~1.70}  & { ~-857.12}  & { ~1.71}  & { +104.71}  & { 5.72} & { -787.37}  & { 2.88}  & { 1.71 }  & { 1.05 }  & { 1338}  & { 1323}  & { 3}\tabularnewline
{ {}}                        & { +26.02}  & { ~1.65}  & { ~-848.67}  & { ~1.65}  & { +111.44}  & { 6.92} & { -790.43}  & { 3.51}  & { 1.93 }  & { 1.21 }  & { 1302}  & { 1302}  & { 3}\tabularnewline
{ C/1991 F2 Helin-Lawrence}  & { +12.53}  & { ~1.79}  & { ~-116.77}  & { ~1.79}  & { +14.24}  & { 4.03}  & { ~-93.14}  & { 7.31}  & { 0.83 }  & { 0.78 }  & { 216}   & { 213}   & { 3}\tabularnewline
{ {}}                        & { +14.64}  & { ~2.13}  & { ~-114.67}  & { ~2.13}  & { +24.90}  & { 4.70}  & { ~-89.76}  & { 12.25} & { 0.89 }  & { 0.86 }  & { 212}   & { 212}   & { 3}\tabularnewline
{ C/1993 A1 Mueller}         & { +11.26}  & { ~3.99}  & { ~-508.85}  & { ~3.99}  & { +62.66}  & { 2.25}  & { -408.00}  & { 2.64}  & { 2.42 }  & { 0.98 }  & { 1488}  & { 1489}  & { 2}\tabularnewline
{ {}}                        & { -19.36}  & { ~3.88}  & { ~-539.43}  & { ~3.87}  & { +57.95}  & { 2.74}  & { -406.83}  & { 4.16}  & { 2.79 }  & { 1.15 }  & { 1471}  & { 1471}  & { 2}\tabularnewline
{ C/1993 Q1 Mueller}         & { ~+6.62}  & { ~2.04}  & { ~~-62.43}  & { ~2.04}  & { +12.24}  & { 6.15}  & { -204.56}  & { 23.40} & { 0.97 }  & { 0.92 }  & { 1047}  & { 1041}  & { 3}\tabularnewline
{ {}}                        & { ~-0.49}  & { ~3.25}  & { ~~-69.55}  & { ~3.27}  & { ~-5.33}  & { 8.70}  & { -222.77}  & { 37.48} & { 1.21 }  & { 1.19 }  & { 1039}  & { 1039}  & { 3}\tabularnewline
{ C/1996 E1 NEAT}            & { -36.48}  & { ~3.68}  & { ~+361.78}  & { ~3.67}  & { +30.15}  & { 4.56}  & { +368.83}  & { 5.44}  & { 0.93 }  & { 0.60 }  & { 495}   & { 492}   & { 2}\tabularnewline
{ {}}                        & { -43.35}  & { ~4.05}  & { ~+354.93}  & { ~4.04}  & { +18.42}  & { 6.50}  & { +374.29}  & { 7.89}  & { 1.01 }  & { 0.77 }  & { 484}   & { 484}   & { 2}\tabularnewline
{ C/1997 BA$_{6}$ Spacewatch}& { ~+1.49}  & { ~0.35}  & { ~+371.77}  & { ~0.35}  & { +31.83}  & { 1.15}  & { +402.48}  & { 1.72}  & { 0.74 }  & { 0.67 }  & { 1054}  & { 1054}  & { 3}\tabularnewline
{ {}}                        & { ~-1.49}  & { ~0.46}  & { ~+369.86}  & { ~0.46}  & { +38.42}  & { 1.68}  & { +405.01}  & { 1.62}  & { 1.17 }  & { 0.89 }  & { 1043}  & { 1043}  & { 3}\tabularnewline
{ C/1997 J2 Meunier-Dupouy}  & { +38.83}  & { ~0.57}  & { ~~~-2.72}  & { ~0.57}  & { +44.64}  & { 0.88}  & { ~+14.72}  & { 0.91}  & { 0.67 }  & { 0.53 }  & { 2881}  & { 2863}  & { 2}\tabularnewline
{ {}}                        & { +36.86}  & { ~0.57}  & { ~~~-4.69}  & { ~0.57}  & { +45.81}  & { 1.03}  & { ~+15.52}  & { 1.00}  & { 0.74 }  & { 0.60 }  & { 2824}  & { 2824}  & { 2}\tabularnewline
{ C/1999 Y1 LINEAR}          & { +42.92}  & { ~0.88}  & { ~+350.42}  & { ~0.88}  & { +47.35}  & { 0.94}  & { +345.69}  & { 1.46}  & { 0.61 }  & { 0.48 }  & { 1747}  & { 1749}  & { 3}\tabularnewline
{ {}}                        & { +19.44}  & { ~0.78}  & { ~+326.90}  & { ~0.78}  & { +49.40}  & { 1.58}  & { +342.34}  & { 1.95}  & { 0.92 }  & { 0.76 }  & { 1737}  & { 1737}  & { 3}\tabularnewline
{ C/2000 SV$_{74}$ LINEAR}   & { +50.23}  & { ~0.40}  & { ~~-85.55}  & { ~0.40}  & { +92.31}  & { 0.85}  & { ~-54.88}  & { 0.60}  & { 1.11 }  & { 0.71 }  & { 4389}  & { 4349}  & { 3}\tabularnewline
{ {}}                        & { +46.66}  & { ~0.39}  & { ~~-89.12}  & { ~0.39}  & { +94.77}  & { 1.18}  & { ~-60.83}  & { 0.75}  & { 1.23 }  & { 0.94 }  & { 4356}  & { 4356}  & { 3}\tabularnewline
{ C/2001 Q4 NEAT}            & { +12.51}  & { ~0.43}  & { ~-731.74}  & { ~0.43}  & { +60.87}  & { 0.48}  & { -696.62}  & { 0.48}  & { 1.29 }  & { 0.63 }  & { 5305}  & { 5263}  & { 3}\tabularnewline
{ {}}                        & { +05.80}  & { ~0.44}  & { ~-738.46}  & { ~0.44}  & { +59.59}  & { 0.70}  & { -696.74}  & { 0.64}  & { 1.69 }  & { 0.80 }  & { 5223}  & { 5223}  & { 3}\tabularnewline
{ C/2002 E2 Snyder-Murak.}   & { +42.95}  & { ~2.31}  & { ~-425.73}  & { ~2.31}  & { +89.59}  & { 12.39} & { -441.69}  & { 2.79}  & { 0.61 }  & { 0.57 }  & { 1863}  & { 1863}  & { 2}\tabularnewline
{ {}}                        & { +38.11}  & { ~2.79}  & { ~-430.57}  & { ~2.79}  & { +75.00}  & { 18.32} & { -445.15}  & { 3.80}  & { 0.77 }  & { 0.75 }  & { 1775}  & { 1775}  & { 2}\tabularnewline
{ C/2002 T7 LINEAR}          & { -13.84}  & { ~0.16}  & { ~-653.57}  & { ~0.16}  & { +20.72}  & { 0.39}  & { -660.08}  & { 1.05}  & { 0.61 }  & { 0.58 }  & { 8596}  & { 8768}  & { 3}\tabularnewline
{ {}}                        & { -24.42}  & { ~0.24}  & { ~-664.14}  & { ~0.24}  & { +19.73}  & { 0.39}  & { -650.86}  & { 1.03}  & { 1.21 }  & { 0.64 }  & { 8643}  & { 8643}  & { 3}\tabularnewline
{ C/2003 K4 LINEAR}          & { +32.08}  & { ~0.16}  & { ~-186.23}  & { ~0.16}  & { +30.73}  & { 0.54}  & { -126.41}  & { 0.81}  & { 0.62 }  & { 0.54 }  & { 7219}  & { 7233}  & { 3}\tabularnewline
{ {}}                        & { +40.36}  & { ~0.17}  & { ~-177.95}  & { ~0.17}  & { +34.73}  & { 0.63}  & { -130.71}  & { 0.77}  & { 0.78 }  & { 0.63 }  & { 7114}  & { 7114}  & { 3}\tabularnewline
{ C/2004 B1 LINEAR}          & { +34.64}  & { ~0.20}  & { ~-481.63}  & { ~0.20}  & { +35.55}  & { 0.72}  & { -460.98}  & { 0.68}  & { 0.53}   & { 0.43}   & { 3777}  & { 3758}  & { 3}\tabularnewline
{ {}}                        & { +34.66}  & { ~0.22}  & { ~-481.61}  & { ~0.22}  & { +38.01}  & { 0.81}  & { -460.07}  & { 0.81}  & { 0.68}   & { 0.58}   & { 3783}  & { 3783}  & { 3}\tabularnewline
\hline
\end{tabular}
}
\end{table*}
\end{center}

\section{Obtaining the most probable original and future orbits}\label{sec:Obtaining-the-most}

To investigate the past and future history of each particular comet
we have examined the evolution of thousands of virtual comets (VCs)
orbits from the confidence region, i.e. the 6-dimensional region
(7--9D region in the NG~case) of orbital elements (and
NG~parameters), each time fully compatible with the observations. We
constructed the confidence region using the Sitarski's method of the
random orbit selection \citep{sitarski:1998}.

This method allows us to generate any number of randomly selected
orbits of VC. It should be noticed that according to this random
selection method the derived sample of VCs follows the normal
distribution in the orbital elements space (also the rms's fulfil
the 6--9~dimensional normal statistics;
\citealp{krolikowska-sit-soltan:2009}).

\begin{figure}
\begin{centering}
\includegraphics[width=4.1cm]{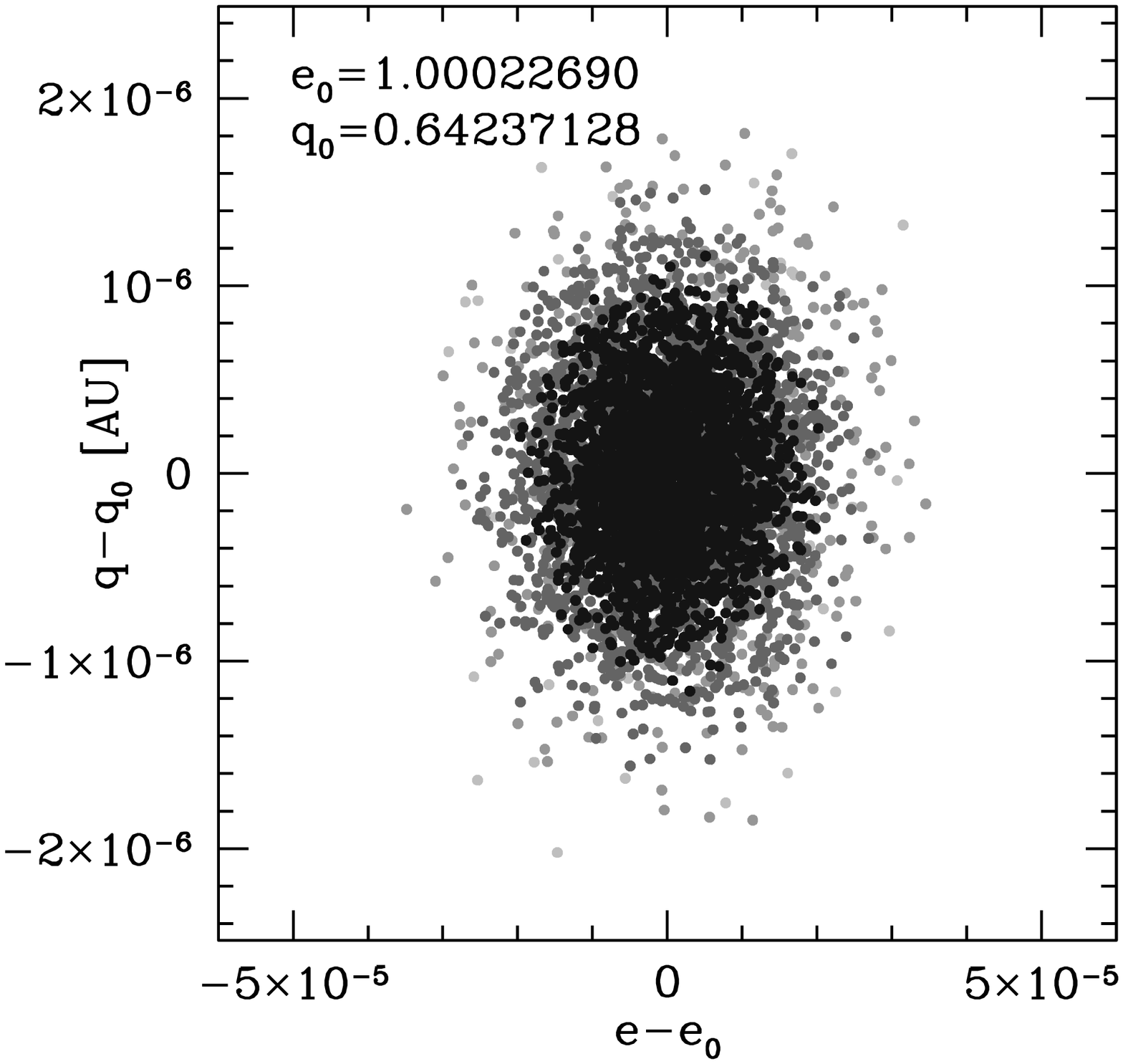} \includegraphics[width=4.1cm]{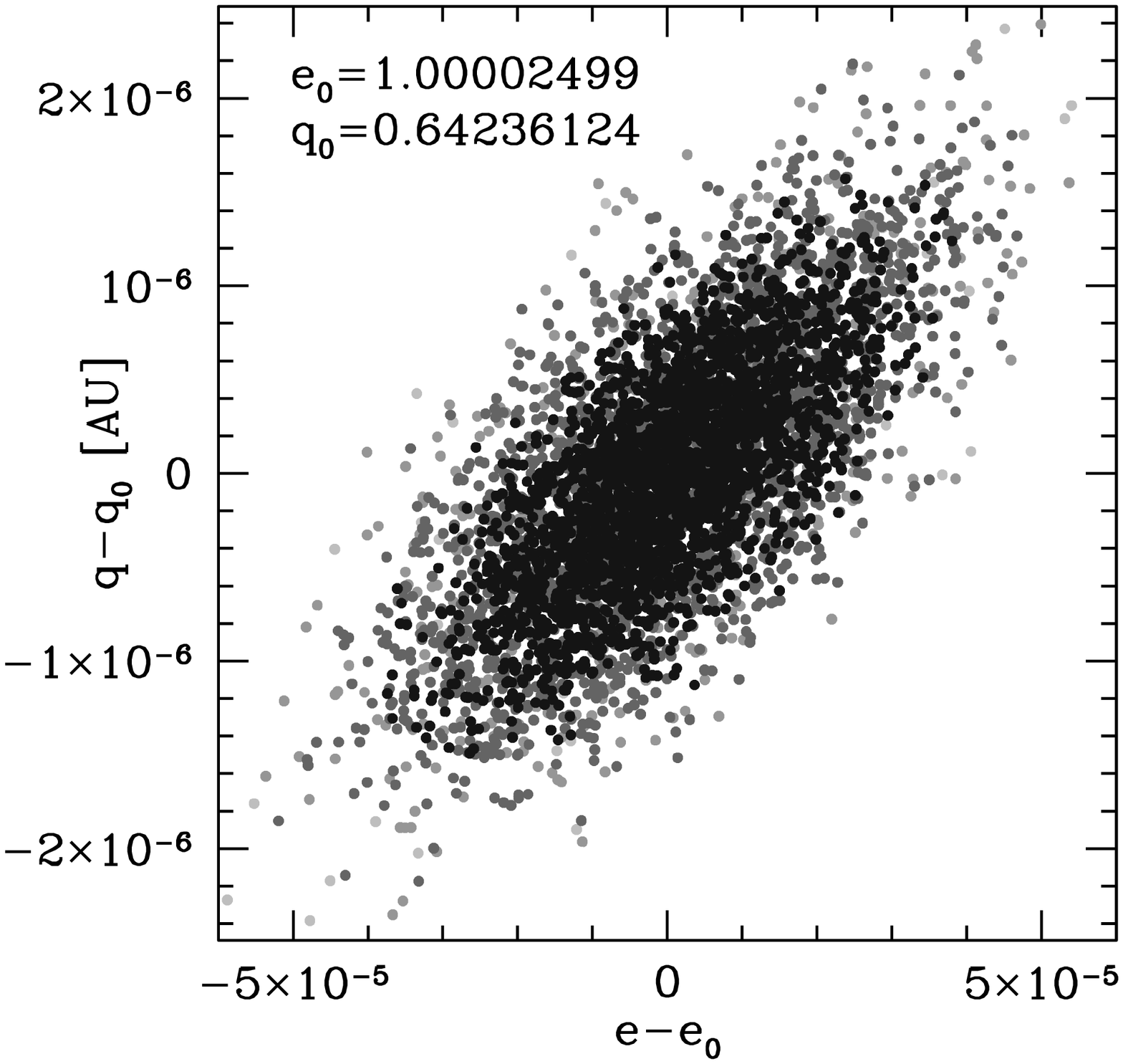}
\includegraphics[width=4.1cm]{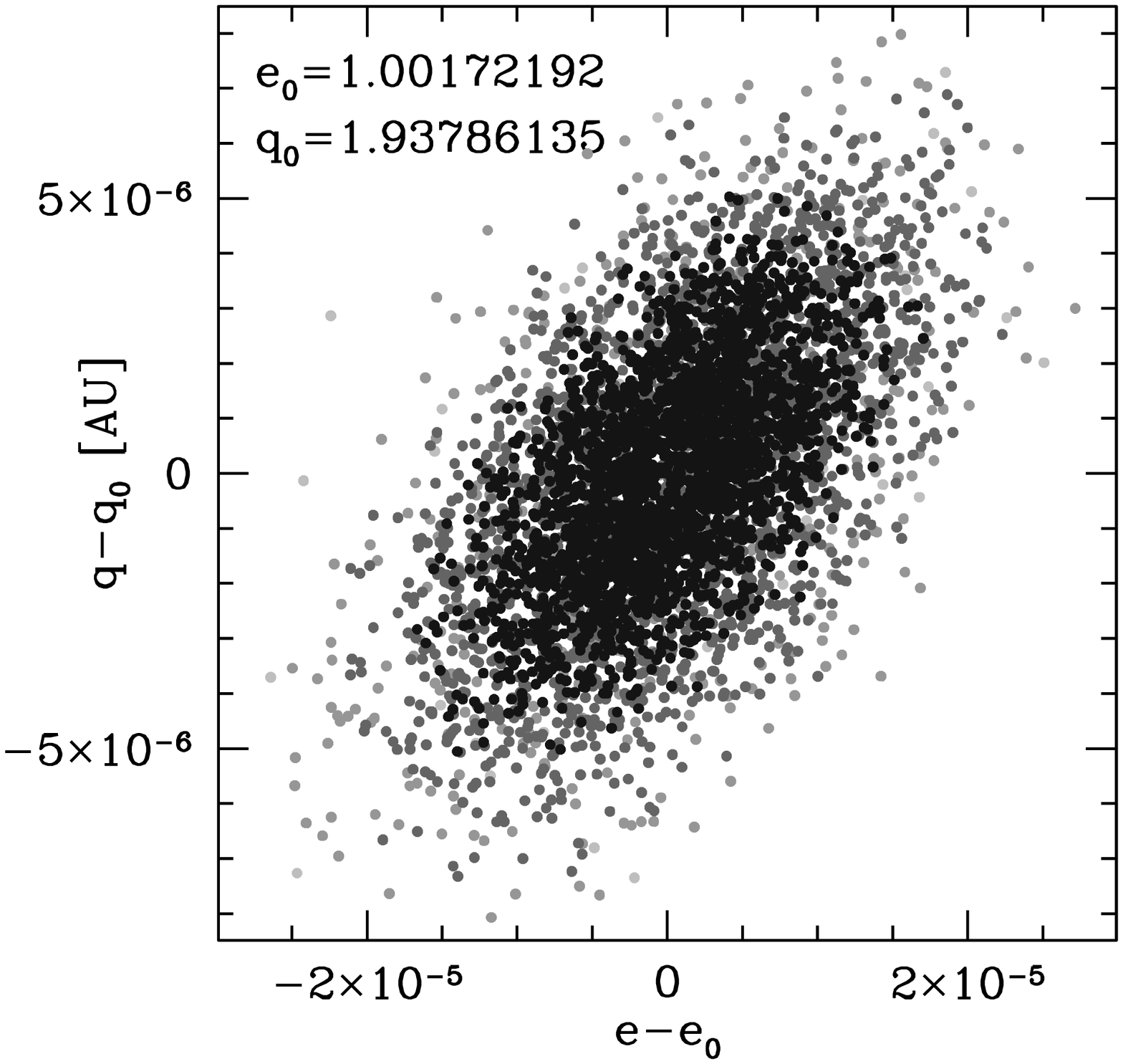} \includegraphics[width=4.1cm]{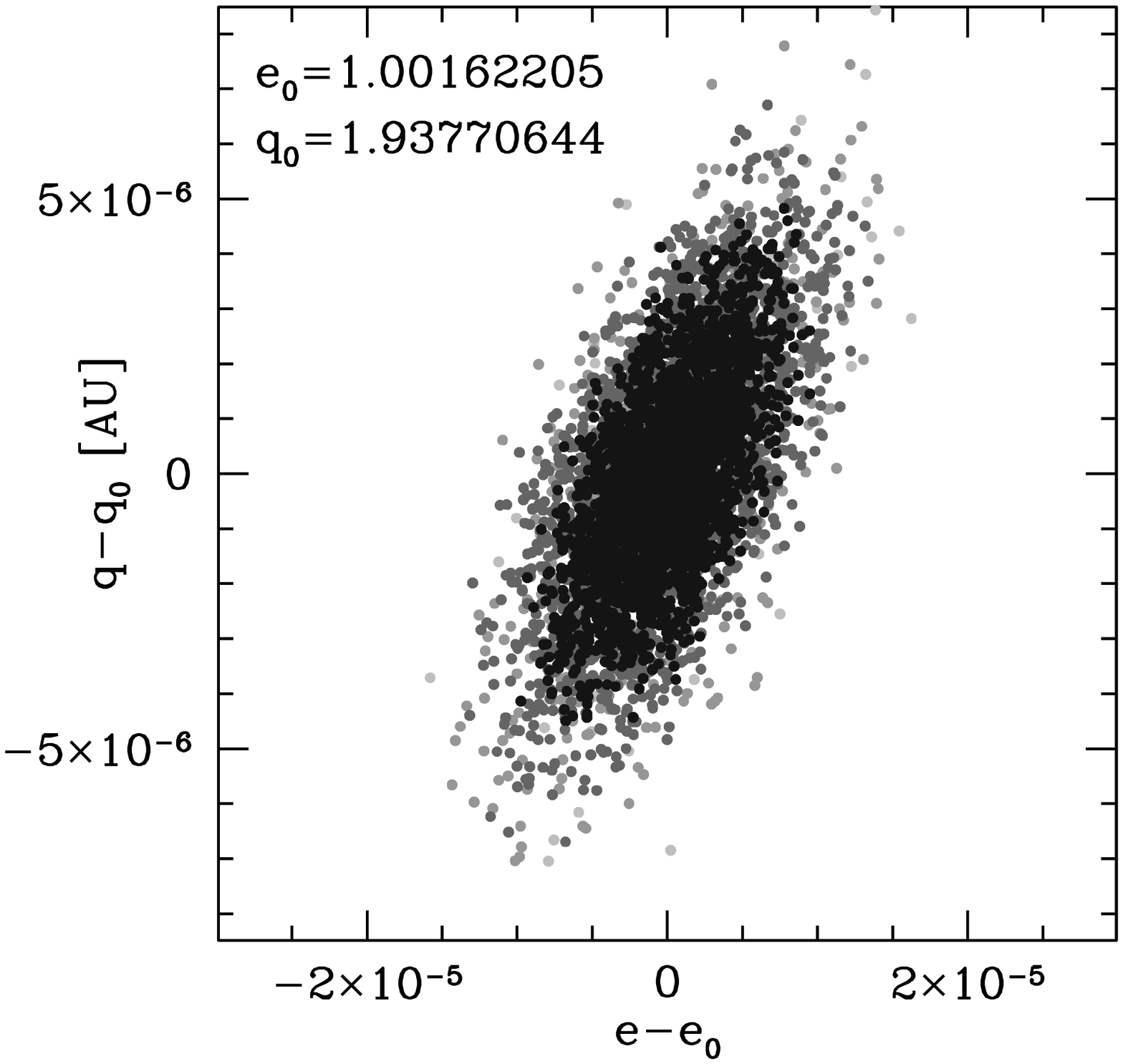}
\par\end{centering}

\caption{Projection of the 6D/8D space of 5\,000~VCs of C/1989~Q1
(upper panel) and C/1993~A1 (lower panel) onto the e-q plane. Left
plots show the pure gravitational swarms of VCs, right plots -- NG
swarms. Each point represents a single virtual orbit, while its grey
tint indicates the deviation magnitude from the nominal orbit with
the confidence level of: $<50$\,\%, $50$\,\% -- $90$\,\%, $90$\,\%
-- $99$\,\%, and $>99$\,\% (from the most dark grey points to the
most light grey points, respectively). Each plot is centered on the
nominal values of respective pair of osculating orbital elements
denoted by the subscript '0'.}

\begin{centering}
\label{fig:93a1_swarm}
\par\end{centering}
\end{figure}

In the calculations described here we fill a confidence region with
5\,000~VCs for each nominal solution (GR and NG) for each considered
comet. Fig.~\ref{fig:93a1_swarm} shows projections of the 6D/8D
parameter space of 5\,000 VCs of C/1989~X1 (upper panels) and
C/1993~A1 (lower panels) onto the plane of two chosen orbital
elements. Orbital cloning procedure was applied close to the
observational arc (at the epoch of 1990\,April\,19 and
1992\,October\,25, respectively). The derived swarm of VCs follows
the normal distribution in the 6D space of orbital elements (8D
space of orbital elements and two NG~parameters in the NG~case for
both comets). This is visualized by four grey tints of points in
Fig.~\ref{fig:93a1_swarm}. Each point represents a single VC, while
its grey tint indicates the deviation magnitude from the nominal
orbit with the confidence levels defined as follows: $<50$\,\%,
$50$\,\% -- $90$\,\%, $90$\,\% -- $99$\,\%, and $>99$\,\% (from the
darkest to brightest points, respectively). The symbols in crowded
areas heavily overlap and the brighter points are often overprinted
with darker ones. One can see that the NG~swarm of C/1989~X1 is more
disperse than its pure gravitational swarm (upper panel of
Fig.~\ref{fig:93a1_swarm}). Similar situation took place in almost
all investigated comets except C/1993~A1 and C/2000~SV$_{74}$. The
largest dispersion of original NG~swarm within our sample of NG~Oort
spike comets appeared in case of comet C/1959~Y1~Burnham where
four~NG~parameters were determined. In the C/1993~A1 case the
compactness of NG~swarm of VCs is higher than compactness of
GR~swarm though in the first case we determined two more parameters
from the same data set than in the latter case (lower panel of
Fig.~\ref{fig:93a1_swarm}). It is a consequence of radically better
NG~orbit fitting to data (rms=0\farcs 98) in comparison to ~pure
gravitational orbit fitting (rms=2\farcs 49). In the majority of
cases, however, the relative decreasing of rms is smaller and the
NG~swarm of VCs is more disperse than pure gravitational swarm, as
expected when increasing parameter space dimensions.

\begin{figure}
\begin{centering}
\includegraphics[clip,width=8.6cm]{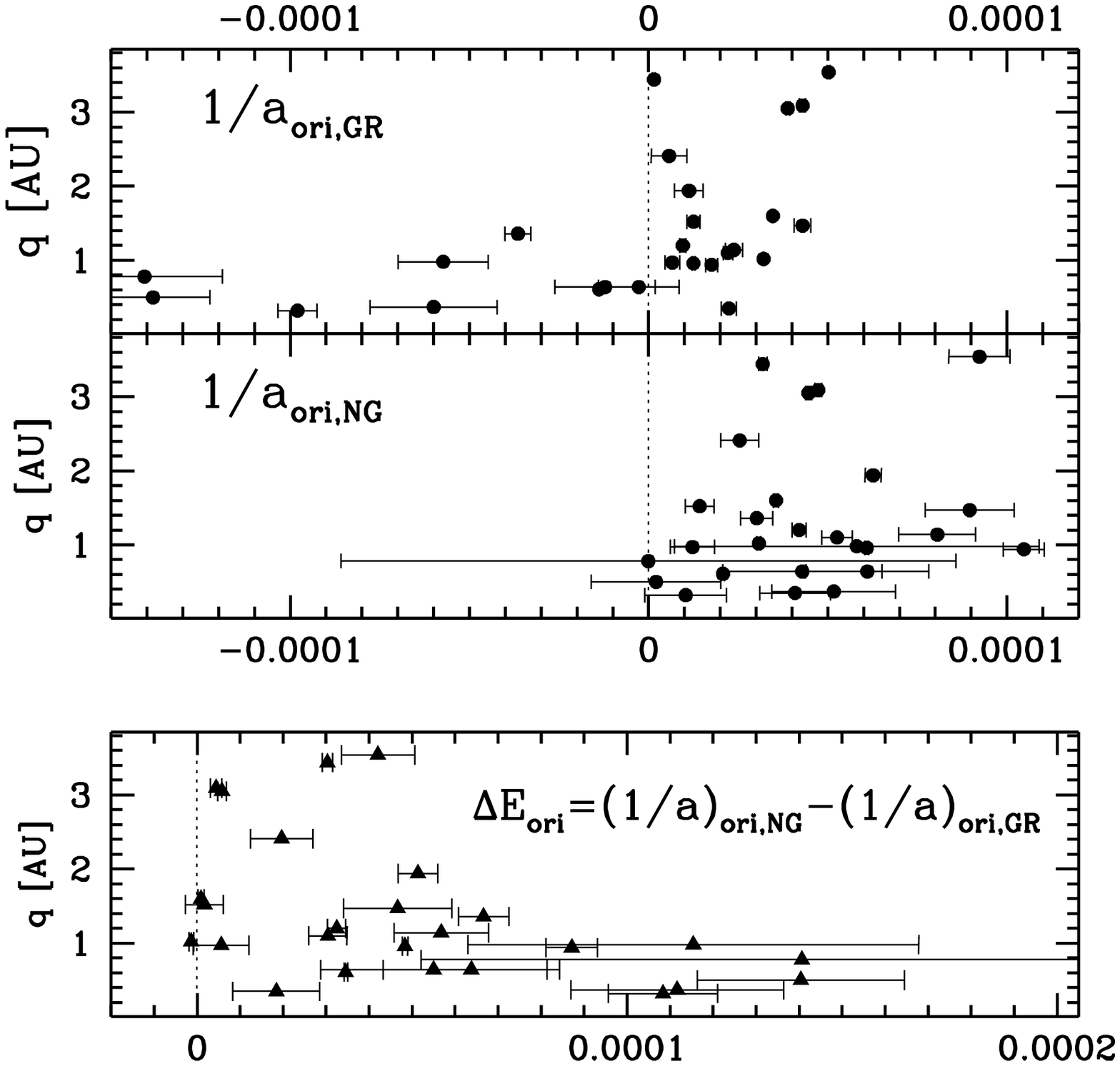}
\par\end{centering}

\caption{Shifts of $1/{\rm a}_{{\rm ori}}$ due to the NG~effects for
26 of investigated comets. Three largest uncertainties of $1/{\rm
a}_{{\rm ori,NG}}$ belong to comets C/1959~Y1 and C/1952~W1 and
C/1892~Q1 (see also Table~\ref{tab:a_original} and
section~\ref{sub:comets_discovered_before}).}

\begin{centering}
\label{fig:spik_ng}
\par\end{centering}
\end{figure}

To calculate the original and future swarm of cometary orbit each of
VC was followed from its position at osculation epoch backwards and
forwards until the VC reached a distance of 250\,AU from the Sun.
The equations of comet's motion have been integrated numerically
using the recurrent power series method
\citep{sitarski:1989,sitarski:2002}, taking into account
perturbations by all the planets and including the relativistic
effects. In this way we were able to obtain the nominal
original/future orbit of each comet as well as the uncertainties of
the derived values of orbital elements by fitting the normal
distribution to each original/future cometary swarm
\citep{krolikowska:2001}. Original and future semi-major axes with
their uncertainties are given in Table~\ref{tab:a_original} where
comets are ordered by discovery date. Fig.~\ref{fig:spik_ng}
visualizes the differences between original $1/{\rm a}$ derived for
NG and GR swarms of comets. One can note that all investigated
comets have $1/a_{{\rm ori,GR}}<5\times10^{-5}$~AU$^{-1}$. This
means, that all these comets would belong to the first row of Table
1 in the \citet{oort:1950} paper.

%

\section{Past and future motion under the Galactic tide}{\label{sec:Past-and-future}}

\subsection{Force model}\label{sub:Force-model}

\begin{figure}
\includegraphics[angle=270,width=1\columnwidth]{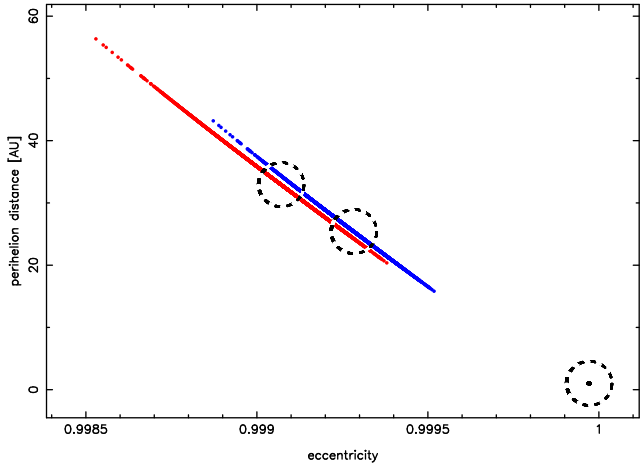}

\caption{\label{fig:galactic_models_1}The comparison of the past motion of
C/2003 K4 for different force-models. The upper (blue) line depicts
the swarm of VCs stopped at previous perihelion when full Galactic
tide (disc + centre) was used. The lower (red) line is for simplified
(disc only) model. For the comparison the starting swarm (very concentrated
in this scale) is shown in the right,lower corner with black dots.
For each swarm the nominal comet orbit is situated at the centre of
the corresponding dashed-line circle.}

\end{figure}

\begin{figure}
\includegraphics[clip,angle=270,width=1\columnwidth]{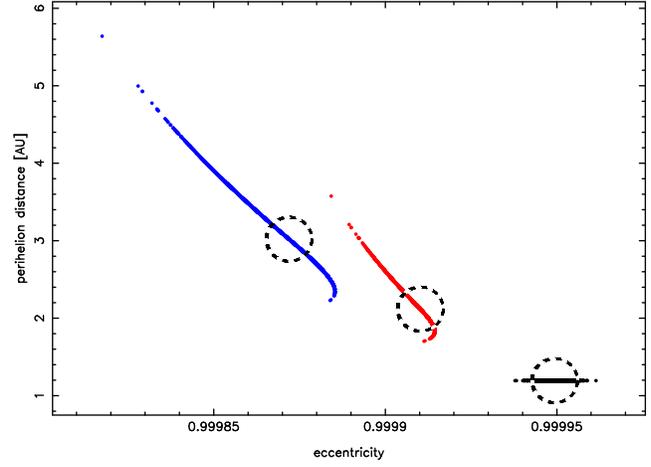}

\caption{\label{fig:galactic_models_2}The comparison of the past motion of
C/1986 P1 for different force-models. All symbols have the same meaning
as in Fig.\ref{fig:galactic_models_1}.}

\end{figure}

Following \citet{fouchard-froeschle-matese-valsecchi:2005} and others,
we used a set of equations of motion in the vicinity of the Sun, accounting
for Galactic disk tide and Galactic centre tide.
\citeauthor{pretka:1998} (\citeyear{pretka:1998}, \citeyear{pretka:1999})
argued that for relatively small time intervals (say ten million years)
it is not necessary to include the Galactic centre term, as its influence
manifests significantly on considerably longer time intervals. However,
to maintain compatibility with other works in the field and taking
into account, that including Galactic centre term into the model of
Galactic tides is not very time consuming we decided to use such a
full model. During all integrations we performed calculations in both:
simplified and full Galactic tide models, just to confirm the above-mentioned
conclusions and observe the difference in some specific cases. It
appeared, that while qualitative evolution is almost the same, the
quantitative results may differ. Some examples of the significant
differences in the output from these two models are given in Figs.\ref{fig:galactic_models_1}
and \ref{fig:galactic_models_2}.

We use here two different frames. The rotating frame is defined as a
heliocentric frame with the $\hat{x}'$ axis in the radial direction
pointing toward the Galactic center, the $\hat{y}'$ axis pointing
along the Sun’s velocity, and $\hat{z}'$ completing a right-handed
system. The nonrotating frame $Oxyz$ coincides with this rotating
one at $t=0$, but keeps its direction fixed. Basing on
\citet{heis-trem:1986} one can write the formula for the tidal force
from the disk and the centre of the Galaxy in the rotating galactic
frame $Ox^{'}y^{'}z^{'}$ as follows\[
\overrightarrow{F}=\frac{GM_{B}}{r^{3}}\overrightarrow{r}+(A-B)(3A+B)x'\hat{x}'+\]
 \begin{eqnarray*}
 &  & -(A-B)^{2}y'\hat{y}'-[4\pi G\rho_{o}-2(B^{2}-A^{2})]z'\hat{z}'\end{eqnarray*}

where $A$ and $B$ are the Oort galactic constants, $G$ is the gravity
constant, $r$ is the heliocentric distance, $\rho_{o}$ is the local
disk matter density and $M_{B}$ is the central mass (the Sun and
planets), expressed in solar masses. Let $G_{1}$, $G_{2}$ and $G_{3}$
be such that:

\[
G_{1}=-(A-B)(3A+B)\]

\[
G_{2}=(A-B)^{2}\]

\[
G_{3}=4\pi G\rho_{o}-2(B^{2}-A^{2})\]

\noindent then the equations of motion in rectangular coordinates
($Oxyz$ inertial frame) are:

\[
\frac{d^{2}x}{dt^{2}}=-\frac{GM_{B}}{r^{3}}x-G_{1}x'\cos(\Omega_{o}t)+G_{2}y'\sin(\Omega_{o}t)\]

\[
\frac{d^{2}y}{dt^{2}}=-\frac{GM_{B}}{r^{3}}y-G_{1}x'\sin(\Omega_{o}t)-G_{2}y'\cos(\Omega_{o}t)\]

\[
\frac{d^{2}z}{dt^{2}}=-\frac{GM_{B}}{r^{3}}z-G_{3}z\]

\noindent where $\Omega_{o}$ is the angular velocity of the Sun
about the Galactic center and as a consequence the angular velocity
of the rotating frame with respect to the nonrotating one and

\noindent  $x'=x\cos(\Omega_{o}t)+y\sin(\Omega_{o}t)$ ~~~~

\noindent $y'=-x\sin(\Omega_{o}t)+y\cos(\Omega_{o}t)$.

\begin{figure}
\includegraphics[clip,angle=270,width=1\columnwidth]{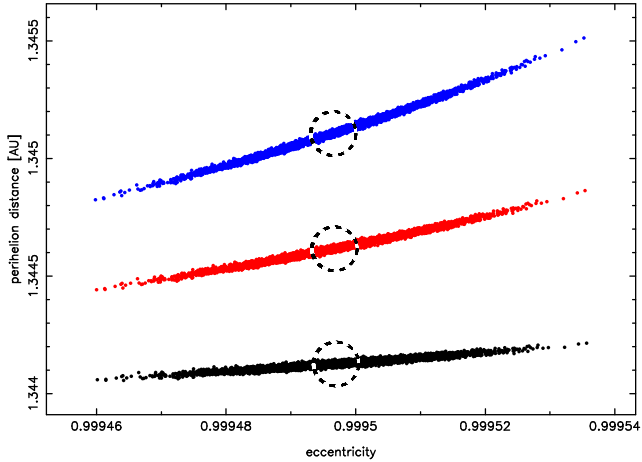}

\caption{\label{fig:diff_rho}The comparison of the future motion of C/1996
E1 with three different $\rho_{o}$ values used. The full Galactic
tide model was used here and all VCs were stopped at 120\,000 AU,
going out of the Solar System. The upper (blue) swarm was obtained
for$\rho_{o}=0.050$~$M_{\odot}pc^{-3}$, the middle one (red) was
obtained for nominal value ( $\rho_{o}=0.100$ ~$M_{\odot}pc^{-3}$)
and the lower swarm (black one) for $\rho_{o}=0.150$~$M_{\odot}pc^{-3}$.
The centre of the dashed-line circle drawn on top of each of the swarms
denotes the position of the nominal comet orbit. }

\end{figure}

\noindent Substituting this into the equations of motion and
remembering that $G_{2}=-G_{1}$ (all contemporary models of Galactic
tides adopt $B^{2}=A^{2}$, see
\citet{fouchard-froeschle-matese-valsecchi:2005} as the most recent
review of the different tide models and their calculation methods)
we obtain:

\[
\frac{d^{2}x}{dt^{2}}=-\frac{GM_{B}}{r^{3}}x-G_{1}x\cos(2\Omega_{o}t)-G_{1}y\sin(2\Omega_{o}t)\]

\[
\frac{d^{2}y}{dt^{2}}=-\frac{GM_{B}}{r^{3}}y-G_{1}x\sin(2\Omega_{o}t)+G_{1}y\cos(2\Omega_{o}t)\]

\[
\frac{d^{2}z}{dt^{2}}=-\frac{GM_{B}}{r^{3}}z-G_{3}z\]

\noindent which is the set we integrate numerically. We used
$G_{1}=-7.0702403\times10^{-10}$, and $G_{3}=4\pi G\rho_{o}$, where
$\rho_{o}$, is the local Galactic mater density in solar masses per
cubic parsec. The model limited to the disk tide can be obtained by
putting $G_{1}=0$, and we used it in parallel, to monitor the
separate influence of the Galactic centre term. Two examples of such
a comparison of different dynamical models are presented in
Figs.~\ref{fig:galactic_models_1} and~\ref{fig:galactic_models_2}.
As it concerns the adopted value of the local Galactic mater density
$\rho_{o}$, we followed most of the contemporary papers in the
field, adopting $\rho_{o}=0.100$~$M_{\odot}pc^{-3}$. To observe the
influence of a possible error in this value we repeated the
calculations for several comets using three significantly different
values: $\rho_{o}=0.050$, $\rho_{o}=0.100$ and
$\rho_{o}=0.150$~$M_{\odot}pc^{-3}$. An example of such a comparison
is presented in Fig.\ref{fig:diff_rho}, which is significantly
magnified to make the differences visible.

\citet{dyb-hab3:2006} have shown, that none of the known stars could
significantly influence the cometary motion recently (during last
10 million years, or so) and in similar time interval in the future.
This time interval is comparable with orbital period of an elliptic
comet with the semi-major axis of 50\,000 AU and thus it is close
to the timescale of the longest past or future cometary motions analyzed
in this paper. \citet{dyb-hab3:2006} searched for real stellar perturbers
in all available catalogues, selected several stars with maximum possible
influence on the Oort cloud comets and showed that comparing with
Galactic perturbations their influence is negligible. He also presented
a discussion on the completeness of the stellar data (besides positions,
both parallax and radial velocity are necessary). In short: we cannot
be sure that some stars passed (or will pass) in the vicinity of the
Sun during 20 million years time interval, centred at present. But
these stars are absent in our catalogues either because they are too
small (e.g. brown dwarf, see for example: \citealp{RECONS}) or move
very fast, and now are too far, or both. But only relatively massive
stars (of order of one solar mass or more) and moving rather slowly
(less than 30 km s$^{-1}$) can significantly change cometary orbit,
even with the aphelion at 100\,000 AU. For more detailed discussion
see \citet{dyb-hab3:2006} .

For the reasons explained above, and taking into account, that precise
calculation of (even very weak) stellar perturbations would be time
consuming, we decided to omit them in our model. Perhaps in future,
especially if some important star will be discovered (and its parallax
and radial velocity will be measured, see \citealp{dyb-kwiat:2003})
we will include its gravitational influence on the Sun and a comet
into our model.

\subsection{Numerical integration}\label{subsec:Num-int}

The original and future orbit calculations produced an output in the
form of a set of 5001 (nominal orbit plus 5000 VCs) barycentric
orbits for each comet under consideration. To observe how Galactic
perturbation action varies depending on small initial condition
changes we separately followed numerically all 5001 VCs, in general,
one orbital period to the past and to the future. We used the well
known, widely used and tested routine RA15
\citep{everhart-ra15:1985} with the force model described above
(with the parameter LL=12 to obtain the highest possible accuracy on
64-bit computers). Since the Galactic tide is relatively weak, the
automatically adjusted integration step was generally very large so
the speed of the calculation allowed us to follow individually such
a big number of VCs for each comet.

\begin{figure}
\includegraphics[angle=270,width=1\columnwidth]{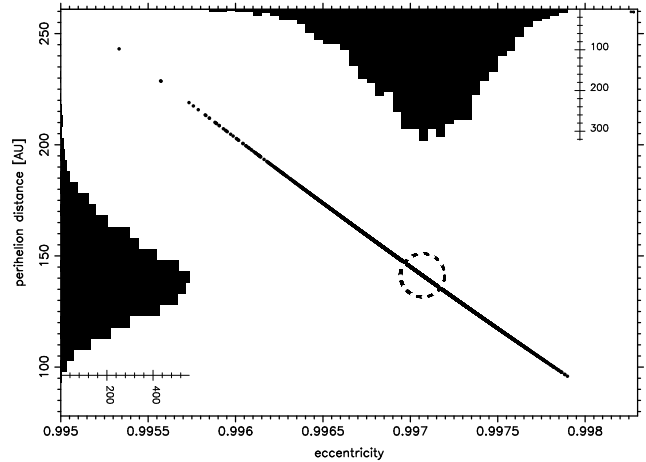}

\caption{\label{fig:past_2002t7}The past motion of C/2002 T7. Presented here
is the distribution in q-e plane of all 5001 VCs for that comet, stopped
at previous perihelion. One can see that the distribution of the previous
perihelion distance is very close to the Gaussian one. In the center
of the dashed-line circle the nominal orbit is located.}

\end{figure}

Since we are interested mainly in the past motion of a comet and the
apparent source of the observed long-period comets we followed their
past motion for one orbital period. Having the opportunity to
investigate also the future motion of these bodies we decided to
perform similar calculations for all of them going one orbital
period to the future. Going further (both back and forth) seems
useless, since a lot of comets have their previous and/or next
perihelion distances deep in the Planetary System. Since there are
no means to calculate precise planetary perturbations for such
bodies say ten million years from now, we cannot follow their motion
{}``behind'' previous and/or next perihelion point. On the other
hand, when a comet have large previous/next perihelion distance,
their subsequent motion do not change our knowledge on the apparent
source of the observed long-period comets, or their future.

\begin{figure}
\includegraphics[angle=270,width=1\columnwidth]{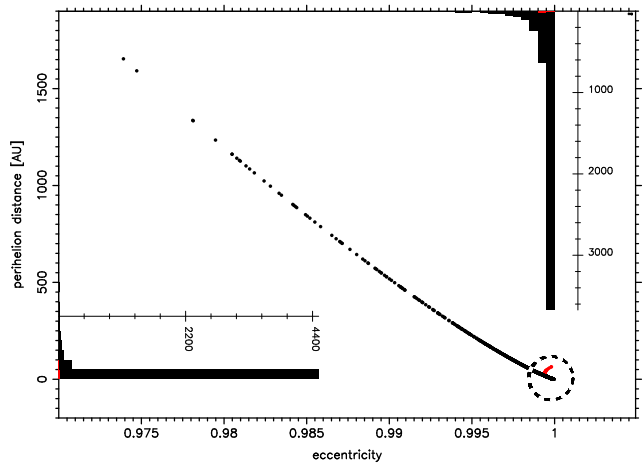}

\caption{\label{fig:past_1989x1asyn}The past motion of C/1989 X1. Presented
here is the distribution in q-e plane of 4974 returning VCs for that
comet (black dots), stopped at previous perihelion and 27 escaping
(while still elliptic) VCs - dark grey (red) dots. One can see that
the distribution of the previous perihelion distance is significantly
different from the Gaussian one. In the center of the dashed-line
circle the nominal (returning in this case) orbit is located. See
text for additional explanation.}

\end{figure}

The situation complicates when past or future orbit of particular VC
is hyperbolic. We have to stop numerical integration at some
heliocentric distance, which we call escape limit. To define this
limit it seems useful to recall contemporary studies on the outer
border of the Oort cloud. Among recent reviews in this field we
mention a chapter in Comets II book by \citet{Dones-W-L-D-book:2004}
and a book by \citet{fernandez_book:2005}. Both present consistent
conclusions that while Oort placed the outer border of the cometary
cloud at 5--15$\times 10^4$\,AU current understanding of cometary
dynamics makes this border much smaller, say  2--8$\times 10^4$\,AU
for the aphelion distance and 1--4$\times 10^4$\,AU for semi-major
axis. Additionally, the majority of the observed Oort spike comets
have the semi-major axes in the range of 20\,000--50\,000 AU for
pure gravitational orbits but this significantly decreases to the
value less than 20\,000\,AU if the NG~forces are included in the
dynamical model. Some discussion of this effect is also in Section
5.2

We decided to be very careful (and a little conservative) so we
decided to adopt the escape limit to be equal 120\,000\,AU. This is
the maximum heliocentric distance (or maximum aphelion distance) at
which we still follow every VCs motion. This was also applied for
very elongated elliptic orbits -- if the particular VC goes further
than 120\,000\,AU from the Sun we call it escaping and stopped the
numerical integration.

This particular choice of the escaping border was unfortunate for a
handful of comets, where small number of VCs traveled a little bit
further than 120\,000 AU, but then return to the Sun. We examined
closely all such cases and if the necessary increase in the outer
limit seemed acceptable we changed it to obtain homogeneous swarm of
returning VCs for that particular comet. Such a procedure was
applied for example for C/1996~E1, where increasing the outer limit
up to 140\,000\,AU allowed all 5001\,VCs to return to their previous
perihelion among Solar System planets when calculating its past
motion. To obtain the same effect for C/1885~X1 one should extend
the outer limit up to 260\,000\,AU, what, in the light of arguments
included in both papers quoted above, seems problematic. At such a
large heliocentric distance the comet motion became chaotic, see
Fig.\,1 in \citet{dyb-hist:2001}.

Due to the fact, that some VCs swarms consists of both returning and
escaping  ones, the rules of stopping the integration were sometimes
more complicated: the escaping part of the swarm was stopped at the
120\,000 AU and the rest of VCs was allowed to reach previous/next
perihelion. There were also a synchronous variant of calculation
stoping: for returning swarms all VCs were stopped when the nominal
one reached previous/next perihelion and for escaping or "mixed"
swarms, all VCs were stopped when the fastest escaping VC reached
120\,000\,AU. This was done to obtain more homogeneous swarm to
analyze, where all VCs are stopped at the same time.

\section{Results}\label{sec:Results}

\begin{table*}
\caption{\label{tab:past_motion}The past distributions of the
NG~swarms of VCs in terms of returning {[}R], escaping {[}E],
including hyperbolic {[}H] VC numbers; the rule of discriminating
between {[}R] and {[}E] are explained in
section~\ref{subsec:Num-int}. Aphelion and perihelion distances are
described either by a mean value for the normal distributions, or
three deciles at 10\%, 50\% (i.e. median), and 90\%. In case of
mixed swarms the mean values or deciles of Q and q are given for
returning part of the VCs swarm. For comparison we included the
osculating perihelion distance in the second column. Last two
columns present the value of $1/{\rm a}_{{\rm ori}}$ and the
percentage of VCs, that one can call dynamically new, basing on
previous q statistics. }

\begin{centering}
\begin{tabular}{lccccccr@{$\pm$}rc}
\hline Comet  & q$_{osc}$  & \multicolumn{3}{c}{Number of VCs} & Q &
q  & \multicolumn{2}{c}{1/a$_{{\rm ori}}$} & \% of VCs\tabularnewline
 & AU  & {[}R]  & {[}E]  & {[}H]  & $10^{3}$AU  & AU  & \multicolumn{2}{c}{$10^{-6}$AU$^{-1}$} & dyn. new \tabularnewline
\hline
C/1885 X1  & 0.64  & ~~4984$^{a}$  & 17  & 1  & 24.2 -- 32.7 -- 50.9  & 0.72 -- 0.95 -- 4.3  & 61.0  & 17.2  & 4.6 \tabularnewline
C/1892 Q1  & 0.98  & ~~3960$^{a}$  & 1041  & 645  & 15.7 -- 28.1 -- 68.6  & 1.00 -- 1.20 -- 28.16  & 58.1  & 50.9  & 30.8 \tabularnewline
C/1913 Y1  & 1.10  & 5001  & 0  & 0  & 34.5 -- 38.0 -- 42.5  & 1.87 -- 2.30 -- 3.14  & 52.6  & 4.2  & 0 \tabularnewline
C/1940 R2  & 0.37  & ~~4886$^{a}$  & 115  & 5  & 27.1 -- 38.3 -- 63.1  & 0.46 -- 0.93 -- 13.5  & 51.7  & 17.3  & 11.5 \tabularnewline
C/1946 U1  & 2.41  & ~~4847$^{a}$  & 154  & 0  & 62.1 -- 78.2 -- 102.3  & 19.7 -- 89.7 -- 517  & 25.4  & 5.2  & 94.3 \tabularnewline
C/1952 W1  & 0.78  & 2137  & ~~2864$^{a}$  & 2464  & 13.4 -- 28.9 -- 78.7  & 0.79 -- 1.03 -- 27.0  & -0.1  & 85.8  & 63.2 \tabularnewline
C/1956 R1  & 0.32  & 1463  & ~~3538$^{a}$  & 909  & 62.3 -- 90.1 -- 113.8  & 0.94 -- 18.0 -- 93.2  & 10.4  & 11.4  & 86.3 \tabularnewline
C/1959 Y1  & 0.50  & 2245  & ~~2756$^{a}$  & 2498  & 9.3 -- 20.5 -- 62.9  & 0.10 -- 0.47 -- 15.5  & 2.1  & 108.1  & 59.6 \tabularnewline
C/1978 H1  & 1.14  & 5001  & 0  & 0  & 21.2 -- 24.9 -- 29.9  & 0.73 -- 0.91 -- 1.00  & 80.5  & 10.7  & 0 \tabularnewline
C/1986 P1  & 1.20  & 5001  & 0  & 0  & 44.9 -- 47.6 -- 50.7  & 2.67 -- 3.11 -- 3.72  & 42.0  & 2.0  & 0 \tabularnewline
C/1989 Q1  & 0.64  & ~~4411$^{a}$  & 590  & 130  & 27.2 -- 42.3 -- 78.6  & 0.41 -- 0.55 -- 33.4  & 42.9  & 22.2  & 24.4 \tabularnewline
C/1989 X1  & 0.35  & ~~4974$^{a}$  & ~~27$^{b}$  & 0  & 37.5 -- 49.1 -- 70.7  & 1.7 -- 5.9 -- 49.4  & 40.8  & 9.9  & 25.9 \tabularnewline
C/1990 K1  & 0.94  & 5001  & 0  & 0  & 17.9 -- 19.1 -- 20.6  & 0.909 -- 0.915 -- 0.920  & 104.7  & 5.7  & 0 \tabularnewline
C/1991 F2  & 1.52  & ~~2971$^{a}$  & 2030  & 0  & 96.6 -- 121 -- 141  & 214 -- 805 -- 2012  & 14.2  & 4.0  & 100 \tabularnewline
C/1993 A1  & 1.94  & 5001  & 0  & 0  & 30.5 -- 31.9 -- 33.4  & 2.40 -- 2.48 -- 2.59  & 62.7  & 2.3  & 0 \tabularnewline
C/1993 Q1  & 0.97  & 1461  & ~~3540$^{a}$  & 72  & 83.4 -- 106.5 -- 122.7  & 75 -- 396 -- 957  & 12.2  & 6.1  & 99.8 \tabularnewline
C/1996 E1  & 1.36  & ~~5001$^{c}$  & 0  & 0  & 55.7 -- 66.5 -- 82.5  & 2.78 -- 9.72 -- 46.3  & 30.2  & 4.6  & 35.5 \tabularnewline
C/1997 BA$_{6}$  & 3.44  & 5001  & 0  & 0  & 60.1 -- 62.8 -- 65.8  & 15.9 -- 19.5 -- 24.7  & 31.8  & 1.2  & 95.3 \tabularnewline
C/1997 J2  & 3.05  & 5001  & 0  & 0  & 44.8 $\pm$ 0.9  & 2.801 $\pm$ 0.017  & 44.6  & 0.9  & 0 \tabularnewline
C/1999 Y1  & 3.09  & 5001  & 0  & 0  & 42.2 $\pm$ 0.8  & 5.82 -- 6.11 -- 6.46  & 47.3  & 0.9  & 0 \tabularnewline
C/2000 SV$_{74}$  & 3.54  & 5001  & 0  & 0  & 21.7 $\pm$ 0.2  & 3.791 $\pm$ 0.009  & 92.3  & 0.8  & 0 \tabularnewline
C/2001 Q4  & 0.96  & 5001  & 0  & 0  & 32.3 -- 32.9 -- 33.5  & 1.875 $\pm$ 0.057  & 60.9  & 0.5  & 0 \tabularnewline
C/2002 E2  & 1.47  & 5001  & 0  & 0  & 18.9 -- 22.3 -- 27.1  & 1.55 -- 1.62 -- 1.77  & 89.6  & 12.4  & 0 \tabularnewline
C/2002 T7  & 0.61  & 5001  & 0  & 0  & 94.0 -- 96.2 -- 98.6  & 142 $\pm$ 19  & 20.7  & 0.4  & 100 \tabularnewline
C/2003 K4  & 1.02  & 5001  & 0  & 0  & 65.1 $\pm$ 1.1  & 11.7 -- 13.8 -- 16.2  & 30.7  & 0.5  & 25.1 \tabularnewline
C/2004 B1  & 1.60  & 5001  & 0  & 0  & 56.3 $\pm$ 1.1  & 2.49 -- 3.17 -- 4.06  & 35.6  & 0.7  & 0 \tabularnewline
\hline
\end{tabular}
\end{centering}

\vspace{-0mm}
 {$\quad${\scriptsize $a$ -- this part of swarm includes
the nominal orbit, }}
 {$\quad${\scriptsize $b$ -- 12 VCs have 200\,000\,AU $<$Q$<$320\,000\,AU
and 3 VCs - Q between 0.4 - 6.1 $\times 10^6$\,km }}
 {$\quad${\scriptsize $c$ -- we used 140\,000\,AU as the outer
border (just 6 VCs have Q$>$120\,000\,AU).}}

\end{table*}

\begin{table*}
\caption{\label{tab:future_motion}The future distributions of the
NG~swarms of VCs in terms of returning {[}R], escaping {[}E],
including hyperbolic {[}H] VC numbers. Aphelion and perihelion
distances as well as the eccentricity are described either by a mean
value for the normal distributions, or three deciles at 10\%, 50\%,
and 90\%. In case of mixed swarms the mean values or deciles of Q, q
and e are given for returning part of the VCs swarm. For comparison
we included the osculating perihelion distance in the second column.
Last column present the value of $1/{\rm a}_{{\rm ori}}$. }

\renewcommand{\tabcolsep}{1.5mm}
\begin{tabular}{@{}lcccccccr@{$\pm$}r@{}}
\hline
Comet  & q$_{{\rm osc}}$  & \multicolumn{3}{c}{Number of VCs} & eccentricity  & Q  & q  & \multicolumn{2}{c}{1/a$_{{\rm fut}}$}\tabularnewline
 & AU  & {[}R]  & {[}E]  & {[}H]  & at 120\,000 AU  & $10^{3}$AU  & AU  & \multicolumn{2}{c}{$10^{-6}$AU$^{-1}$}\tabularnewline
\hline
C/1885 X1  & 0.64  & 8  & ~~4993$^{a}$& ~~4969$^{a}$& 1.000037-1.000101-1.000206 & -  & -  & -102.3  & 41.1 \tabularnewline
C/1892 Q1  & 0.98  & 0  & 5001  & 5001  & 1.000494$\pm$0.000039        & -  & -  & -486.0  & 23.2 \tabularnewline
C/1913 Y1  & 1.10  & 5001  & 0  & 0     & 0.9999154$\pm$0.0000059      & 21.7-23.0-24.6  & 0.945-0.975-0.997  & 86.8  & 4.2 \tabularnewline
C/1940 R2  & 0.37  & 0  & 5001  & 5001  & 1.000759$\pm$0.000034        & -  & -  & -1494  & 36 \tabularnewline
C/1946 U1  & 2.41  & 5001  & 0  & 0     & 0.999765-0.999808-0.999819   & 34.0-38.3-44.0  & 3.07-3.67-5.17  & 52.2  & 5.3 \tabularnewline
C/1952 W1  & 0.78  & 1509&~~3492$^{a}$&~~3229$^{a}$& 0.9999306-1.0000196-1.0000540& 12.3 - 28.6 - 75.6  & 0.13 - 0.71 - 7.61  & -38.9  & 106.7\tabularnewline
C/1956 R1  & 0.32  & 0  & 5001  & 5001  & 1.000048$\pm$0.000014        & -  &  & -579.7  & 10.0 \tabularnewline
C/1959 Y1  & 0.50  & 0  & 5001  & 5001  & 1.00106$\pm$0.00024          & -  & -  & -286.3  & 33.4 \tabularnewline
C/1978 H1  & 1.14  & 0  & 5001  & 5001  & 1.003062$\pm$0.000041        & -  & -  & -1025  & 8 \tabularnewline
C/1986 P1  & 1.20  & 5001  & 0  & 0     & 0.9990777$\pm$0.0000028      & 2.606$\pm$0.008  & 1.202057$\pm$0.000001  & 767.2  & 2.3 \tabularnewline
C/1989 Q1  & 0.64  & ~~4950$^{a}$& 51& 6& 0.9999031-0.9999336-0.9999402& 17.3-24.9-43.0  & 0.67-0.76-1.97  & 80.6  & 26.5 \tabularnewline
C/1989 X1  & 0.35  & 0  & 5001  & 5001  & 1.000552$\pm$0.000025        & -  & -  & -368.0  & 5.4 \tabularnewline
C/1990 K1  & 0.94  & 0  & 5001  & 5001  & 1.001618$\pm$0.0000013       & -  & -  & -790.4  & 3.5 \tabularnewline
C/1991 F2  & 1.52  & 0  & 5001  & 5001  & 1.000191-1.000258-1.000340   & -  & -  & -92.2  & 7.3 \tabularnewline
C/1993 A1  & 1.94  & 0  & 5001  & 5001  & 1.0003624$\pm$0.0000015      & -  & -  & -407.5  & 2.6 \tabularnewline
C/1993 Q1  & 0.97  & 0  & 5001  & 5001  & 1.000746-1.000974-1.001242   & -  & -  & -204.1  & 23.3 \tabularnewline
C/1996 E1  & 1.36  & 5001  & 0  & 0     & 0.9995040$\pm$0.0000073      & 5.32-5.42-5.53  & 1.34466$\pm$0.00004  & 368.8  & 5.4 \tabularnewline
C/1997 BA$_{6}$  & 3.44 & 5001  & 0 & 0 & 0.9986186$\pm$0.0000059      & 4.966$\pm$0.021  & 3.432091$\pm$0.000006  & 402.5  & 1.7 \tabularnewline
C/1997 J2  & 3.05  & 151& 4850$^{a}$& 0 & 0.9999319$\pm$0.0000032$^{b}$& 124-134-145$^{b}$  & 4.568$\pm$0.063$^{b}$  & 14.9  & 0.9 \tabularnewline
C/1999 Y1  & 3.09  & 5001  & 0  & 0     & 0.9989332$\pm$0.0000045      & 5.783$\pm$0.024  & 3.08617$\pm$0.00003  & 345.7  & 1.5 \tabularnewline
C/2000 SV$_{74}$& 3.54  & 0 & 5001 &5001& 1.0001075$\pm$0.0000036      & -  & -  & -54.4  & 0.6 \tabularnewline
C/2001 Q4  & 0.96  & 0  & 5001  & 5001  & 1.000530$\pm$0.000001        & -  & -  & -695.4  & 0.5 \tabularnewline
C/2002 E2  & 1.47  & 0  & 5001  & 5001  & 1.0005871$\pm$0.0000065      & -  & -  & -441.6  & 2.8 \tabularnewline
C/2002 T7  & 0.61  & 0  & 5001  & 5001  & 1.0015129$\pm$0.0000045      & -  & -  & -660.0  & 1.1 \tabularnewline
C/2003 K4  & 1.02  & 0  & 5001  & 5001  & 1.001200$\pm$0.000015        & -  & -  & -125.1  & 0.8 \tabularnewline
C/2004 B1  & 1.60  & 0  & 5001  & 5001  & 1.002920$\pm$0.000008        & -  & -  & -460.6  & 0.7 \tabularnewline
\hline
\end{tabular}

{$\qquad$ \scriptsize $a$ -- this part of swarm includes the nominal orbit,}
{$\qquad$ \scriptsize $b$ -- in this case statistics are for synchronous swarm (see text). }

\end{table*}

\begin{figure}
\includegraphics[angle=270,width=1\columnwidth]{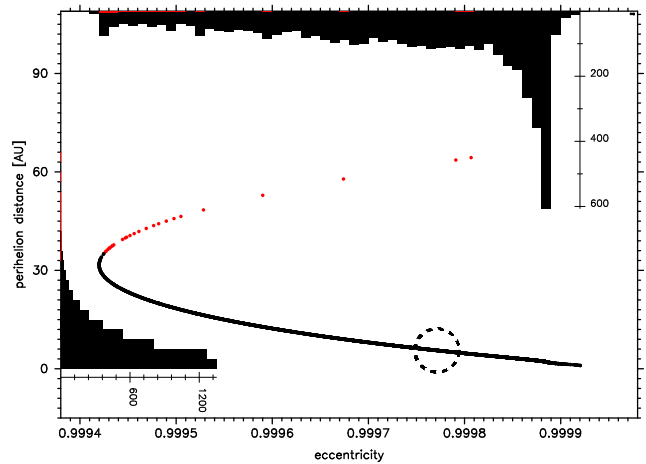}

\caption{\label{fig:past_1989x1syn}The same swarm of VCs for the comet C/1989
X1, but all were stopped synchronously when the fastest escaping VC
reached 120000~AU. Number of returning and escaping VCs is the same
of course (4974 and 27, respectively). See text for additional explanation.}

\end{figure}

\subsection{Overview}\label{sub:Overview}

Let us start with the context, in which our results should be interpreted.
In his paper, \citet{oort:1950} treated original semi-major axes
as definitive for describing past motion of a comet. In the same time
he stated, that all comets with $1/a_{{\rm ori}}<1\times10^{-4}$\,AU$^{-1}$
should be treated as dynamically new, i.e. they are visiting the observable
region for the first time. Such a conclusion came from two assumptions.
First, the only possible perturbation source are stellar perturbations,
what means that going to the past we should follow original, unperturbed
orbit as long as a comet reaches the heliocentric distance large enough
to be perturbed by passing stars. Second, all observed long-period
comets were significantly perturbed by stars (if not, their keplerian,
original orbits bring them exactly to the same perihelion distance
at previous passage and no one can call them {}``dynamically new'').

Today, none of this two assumptions seem to be fully correct. As we
noted in the Introduction, a comet motion now cannot be treated as
unperturbed and the dominating (if not the only one) perturbing force
is the differential, gravitational action of our Galaxy. While Galactic
tides perturb the semi-major axis practically on the negligible level
during one orbital period, the most important part of the past and
future cometary orbit evolution manifests in the eccentricity changes.

As a result, while semi-major axis remains almost unchanged, the
perihelion distance can change significantly, even during one
orbital period both in past and future motion of course. Taking
C/1913~Y1 as typical example, one can note that the original
semi-major axis of this comet a$=$19022.723\,AU and one orbital
period before it was equal 19022.794\,AU. At the same time original
perihelion distance q$=$1.10\,AU and one orbital period earlier it
was equal 2.30\,AU. Small changes in the semi-major axis under the
Galactic perturbations is a well known fact. In the averaged problem
of the disk tide the semi-major axis is even an integral of motion
(see \citet{breiter-dyb:1996} for details). \citeauthor{pretka:1998}
(\citeyear{pretka:1998}, \citeyear{pretka:1999}) showed that slow
secular changes in the semi-major axis due to the action of the
Galactic centre can be observed on the much longer time intervals
than considered here. It means that looking only at 1/a$_{\rm ori}$
of a comet nobody can say what was its previous (or next)
perihelion distance.

\subsection{Past and future evolution - some statistics}\label{sub:past-fut-stat}

As it was described in sections \ref{sec:Observations-and-their},
\ref{sec:Obtaining-the-most} and \ref{sec:Past-and-future}, we very
carefully determined orbits for 26 observed long-period comets using
sophisticated observations treatment and NG force model. Than, at
osculating epoch, we created 5000 VCs for each comet and precisely
propagate all 5001 massless bodies back and forth in time to the
limit of planetary perturbations. \citet{todorov-juch:1981} showed
that the planetary perturbations in the comet's motion should be
included up to 150--200\,AU. Thus, we chose 250\,AU as safely
distant limit where Galactic perturbations are still negligible.
Next, we followed numerically past and future motion of each comet
swarm (consisting of nominal orbit and 5000 VCs) under the influence
of Galactic tides. We stopped the calculation either at past/future
perihelion or when the body in question crosses the assumed
{}``escape'' border, defined at 120\,000~AU.

The results of the investigation of past motion of all 26 comets are
summarized in Table \ref{tab:past_motion}, while the same for future
motion is shown in Table \ref{tab:future_motion}. We present here
only results for NG model as we are convinced that they gives better
description of real cometary orbit and its evolution. For sake of
comparison we prepared extended version of
Tables~\ref{tab:past_motion} and \ref{tab:future_motion} consisting
of GR and NG results and their propagation end-states, but due to
the volume of information, this is available as the on-line material
accompanying this paper only \citep{dyb-krol:2009}.

\begin{figure}
\includegraphics[angle=270,width=1\columnwidth]{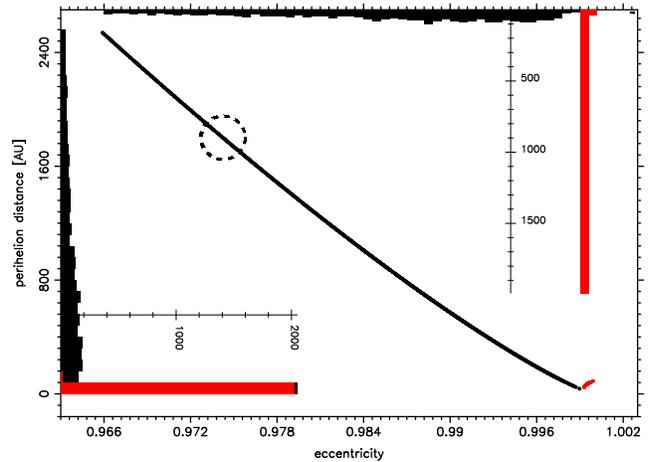}

\caption{\label{fig:past_1991f2asyn}The past swarm of VCs for the comet C/1991~F2.
Here 2971 returning VCs (black dots) were stopped at previous perihelion
and 2030 escaping VCs (dark-gray or red dots) were stopped when crossed
120000~AU. Note that due to the spread distribution of returning
VCs, lots of escaping VCs are overprinted with each other.}

\end{figure}

Table \ref{tab:past_motion} presents the most important result of
this paper. For each comet we show, how many VCs return to previous
perihelion (column {[}R]), how many cross the adopted
\textbf{\emph{escape border}} at 120\,000~AU ({[}E] column) and
additionally how many of escaping VCs have formally hyperbolic
orbits at the end of our calculation ({[}H] column, which presents
subset of {[}E] column). Next three columns are discussed below.
Last column shows the percentage of VCs, that one can call
dynamically new, basing on previous perihelion distance statistics,
i.e. we calculate the part of dynamically new VCs by adding the
number of returning VCs with q$_{\rm prev}>$15\,AU to the number of
escaping VCs. It is remarkable that none of the comets from our
sample can be called for sure hyperbolic before they enter the
sphere of visibility. Three of them, namely C/1982 Q1, C/1952 W1 and
1959 Y1 have significant (but always less than 50~\% of all VCs)
percentage of hyperbolic VCs and only one, C/1952 W1 have formally
negative solution for the inverse semi-major axis:
$1/a=\left( -0.1\pm85.8\right)\times10^{-6}$~AU. However, this value was
determined with the largest relative uncertainty, what results
mainly from the smallest number of observations.

For 15 comets all 5001 VCs are returning and for additional 5 there
is only small (or very small) percentage of escaping VCs,
constituting typical \textbf{\emph{long tails}} of a broad
distribution.

For each of 26 comets investigated here we obtain all or significant
percentage of returning VCs in their past motion. Thus, we were able
to present previous aphelion and perihelion distances statistics for
the returning part of each swarm, see Table~\ref{tab:past_motion}.
In small number of cases when the distribution remains Gaussian we
present its mean value and standard deviation. An illustration of
such a case is given in Fig.~\ref{fig:past_2002t7}. However, in the
majority of cases the nonlinear action of the Galactic tide have
deformed the distributions and they cannot be approximated with the
Gaussian distribution. In such cases we decided to describe these
distributions by three deciles, namely values separating 10\%, 50\%
(median value) and 90\% of all VCs for a particular comet. An
example of the non-Gaussian distribution is given in
Fig.~\ref{fig:past_1989x1asyn}. Here 4974 VCs are returning (black
dots). The strange placement of 27 escaping VCs (dark grey or red
dots) is caused by the fact, that they were stopped at 120\,000~AU
border, while returning VCs were stopped at previous perihelion,
much later. If we stopped all 5001 VCs synchronously, when the
fastest escaping VC reached 120\,000~AU the distribution of VCs in
the q--e plane will be quite different, what can be seen in
Fig.~\ref{fig:past_1989x1syn}.

\begin{figure}
\includegraphics[angle=270,width=1\columnwidth]{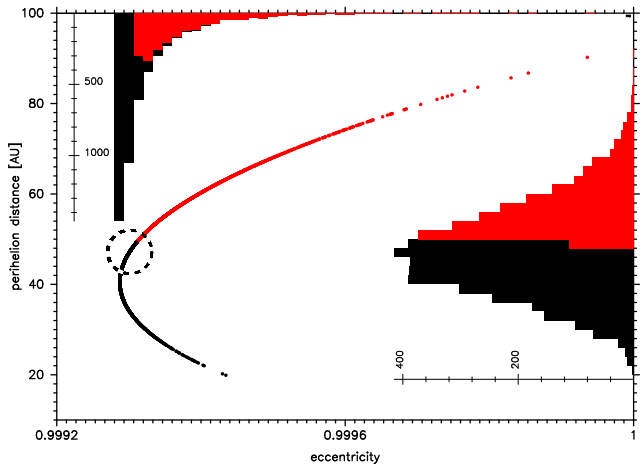}

\caption{\label{fig:past_1991f2syn}The same swarm of VCs for the comet C/1991
F2 as in Fig.\ref{fig:past_1991f2asyn}, but all were stopped synchronously
when crossed 120000~AU.}

\end{figure}

The most complicated cases occur when the swarms of VCs divides into
two distinct parts: returning and escaping (very often with significant
number of hyperbolic past orbits). An example of such a situation
is presented in Figs.~\ref{fig:past_1991f2asyn} and \ref{fig:past_1991f2syn}.

As it was already pointed by
\citet{krolikowska:2001,krolikowska:2004,krolikowska:2006a}, taking
into account NG forces in the procedure of orbit determination shows
that LP~comets generally have smaller original aphelion distances
than widely quoted. A typical previous aphelion distance varies from
20\,000 to 50\,000 AU, what makes the average semi-major axis to be
less than 20\,000 AU. This is a significantly smaller value that
typically thought for so called Oort cloud comets.

However, the most important information about the provenance of
long-period comets is their previous perihelion distance. It is
really striking that 50\% of comets in our sample definitely visited
the Planetary System at their previous perihelion, sometimes as deep
as 1 AU (or less) from the Sun. No one should again call them
\textbf{\emph{dynamically new}} comets! One of the most remarkable
example is comet C/1990~K1, commonly considered as dynamically new
and discussed in detail in Section~\ref{sub:C/1990-K1-Levy}.
Additionally, for the remaining half of our sample, only six comets,
namely C/1946~U1, C/1956~R1, C/1991~F2, C/1993~Q1, C/1997~BA$_{6}$
and C/2002~T7 seem to be dynamically new with high degree of
certainty. Remaining 7 comets have mixed past swarms of VCs, with
the percentage of the previous perihelion distance greater then
15~AU (the widely adopted threshold value for significant planetary
perturbations) varying from 25\% to 63\%.

Performing a very simple (while really impressive) statistics, one
can sum all VCs for all 26 comets and calculate the percentage of
VCs with previous perihelion distance larger than 15~AU.
\textbf{\emph{The result is}} \textbf{\emph{32.9\%.}} Statistically
this is the percentage of dynamically new comets among observed
long-period comets, basing on our sample of 26 comets.

This value is significantly smaller than obtained by
\citeauthor{dyb-hist:2001} (\citeyear{dyb-hist:2001,dyb-hab3:2006}), who claimed that about 50\% of
long-period comets with semi-major axes larger than 10\,000~AU could
be treated as dynamically new. This discrepancy shows the importance
of NG forces applied at the stage of orbit determination
(\citeauthor{dyb-hist:2001} \citeyear{dyb-hist:2001}, \citeyear{dyb-hab3:2006} used pure gravitational
orbits). It is really impressive if we keep in mind that
gravitational original orbit of eight comets from our sample are
hyperbolic and only four of them might be called dynamically new
when NG original orbit is properly accounted for.

\begin{figure}
\includegraphics[angle=270,width=1\columnwidth]{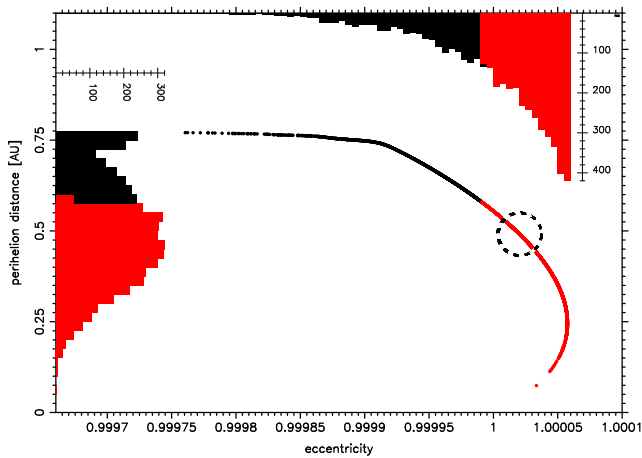}

\caption{\label{fig:fut_1952w1syn}The swarm of VCs for the future motion of
C/1952 stopped synchronously when the fastest escaping VC crossed
120\,000~AU.}

\end{figure}

As it was mentioned earlier, we also studied the future motion of
all 26 comets in our sample, in an analogous manner. The results of
this part of investigation are summarized in Table \ref{tab:future_motion}.
The description is much simpler here, because we have only one comet,
C/1952 W1, with significantly mixed swarm of VCs, consisting of 1509
returning VCs and 3492 escaping (with 3229 hyperbolas among them,
including the nominal orbit). A synchronous plot of all 5001 VCs for
this comet is depicted in Fig.\ref{fig:fut_1952w1syn}. It seems that
this comet will be lost in the future. The similar fate can be attributed
for next 16 comets having 100\% of escaping, hyperbolic VCs and for
C/1885 X1 (over 99\% escaping hyperbolic VCs).

Only 6 comets, C/1913~Y1, C/1946~U1, C/1986~P1, C/1996~E1, C/1997~BA$_{6}$
and C/1999~Y1 have 100\% of returning VCs and all of them will have
the future perihelion distance deep among Solar System planets. The
future motion of these comets seems to be unpredictable due to possible
planetary perturbations.

Two special cases of future motion: C/1989~Q1 and C/1997~J2 are
discussed in the next section.

\begin{figure}
\includegraphics[angle=270,width=1\columnwidth]{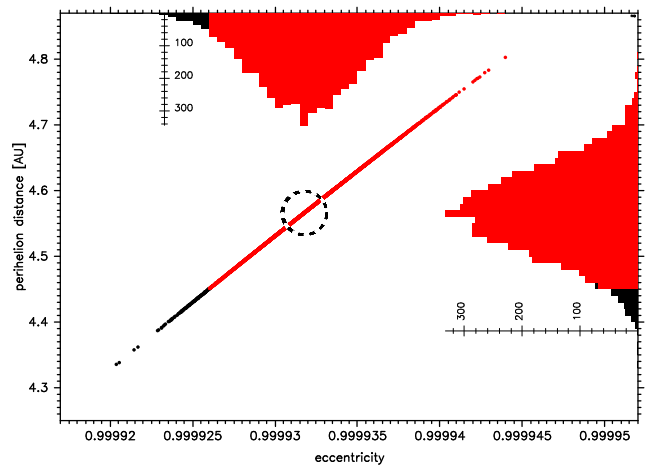}

\caption{\label{fig:fut_1997j2syn}The swarm of VCs for the future motion of
C/1997 J2 stopped synchronously when the fastest escaping VC crossed
120\,000~AU.}

\end{figure}

Summarizing, from our sample of 26~comets 17 should be treated as
lost in the future, one (C/1997 J2) may remain bound to the Solar
System on large elliptical orbit and 8 will return in the future,
pretending they are coming from the so called inner Oort Cloud, and
having perihelion distances well inside the sphere of visibility.

\subsection{Individual cases}\label{sub:Peculiar-cases}

\subsubsection{C/1885 X1 Fabry}\label{sub:C/1885-X1-Fabry}

\begin{figure}
\includegraphics[angle=270,width=1\columnwidth]{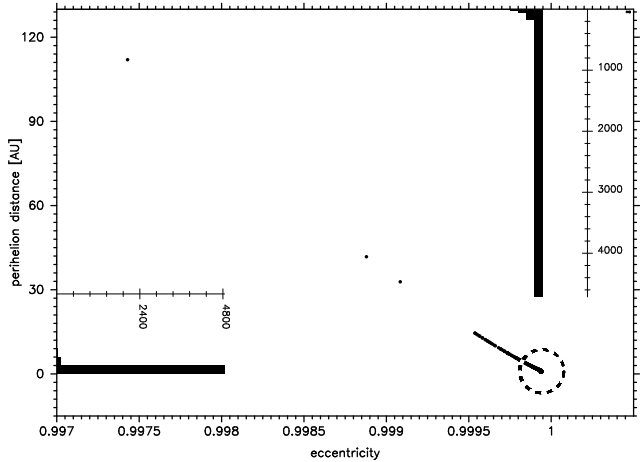}

\caption{\label{fig:1885_A2const}The swarm of VCs for the pas motion of C/1885
X1 stopped at previous perihelion. VCs were generated with the non-gravitational
parameter A$_{2}$kept constant.}

\end{figure}

This is the oldest comet investigated in this paper. Its past motion
is almost clear. When dealing with different $5000$ VC swarms of
this comet, sometimes we obtain just few (or a dozen) of VCs with
hyperbolic past orbit. In every case, there is also some, very small,
percentage of VCs, which have their aphelia slightly above the outer
limit of the heliocentric distance, adopted in this paper as the limit
for returning comet (120\,000 AU). The reason for this is the fact,
that the $A$$_{2}$ non-gravitational parameter is hard to determine
and have relatively large uncertainty ($A$$_{2}=$ $\left( -0.2879\pm0.1606\right)\times10^{-8}$\,AU\,day$^{-2}$
). Together with relatively large mean error ($3.58$ arc-sec, anyway,
very good for such an old comet!) it causes the VC swarms to be more
spread than in many other cases. Assuming, that the $A$$_{2}$ value
is still acceptable, we performed two additional investigations for
this comet, using two non-standard VC swarms: first produced with
fixed $A$$_{2}$ value and the second, with fixed both $A$$_{1}$
and $A$$_{2}$. The result of this additional calculation is very
impressive, what is clearly visible in Fig.\ref{fig:1885_A2const}.
The swarm obtained with fixed $A$$_{2}$ value appeared much more
compact, all VCs are returning in their past motion even for standard
limit of 120\,000 AU. Additional fixing of value $A$$_{1}$ in the
second case did not change the situation significantly (it has much
smaller error of determination). For example, the previous perihelion
distance for standard swarm ($17$ VCs escaping) varied from near
zero to almost 1000 AU (however 90\% of VCs have $q<4.3$AU). For
the swarm obtained with fixed $A$$_{2}$ value all VCs are returning
and have the previous perihelion distance less than $120$ AU, with
only $3$ of them greater than $15$\,AU and 90\% smaller than 1.72\,AU!

Taking all this into account, we can state with high degree of certainty
that this comet was not a dynamically new one, having the $1/a$$_{\rm ori}$
= ($61.0\pm17.2$)$\times10^{-6}$AU$^{-1}$ and the most probable
previous perihelion distance of about 1~--~2\,AU, i.e. deep among
planets. It may be worth to mention here, that this is one of the
famous 19 long-period comets which Oort used to support his hypothesis
in 1950.

The future motion of this comet is much simpler to describe. The vast
majority (99.9\%) of VCs in our standard model are escaping in their
future motion (almost all of these -- 99.3\% of all VCs -- are hyperbolic).
The single, smallest returning orbit has ${\rm q}_{\rm fut}=14$\,AU and
${\rm Q}_{\rm fut}=$60\,000\,AU.
The $1/a$$_{fut}$ = ($-102.0\pm41.1$)$\times10^{-6}$AU$^{-1}$
here, so this comet is almost certainly lost from the Solar System.

\subsubsection{C/1913 Y1 Delavan}\label{sub:C/1913-Y1-Delavan}

\begin{figure}
\includegraphics[width=1\columnwidth]{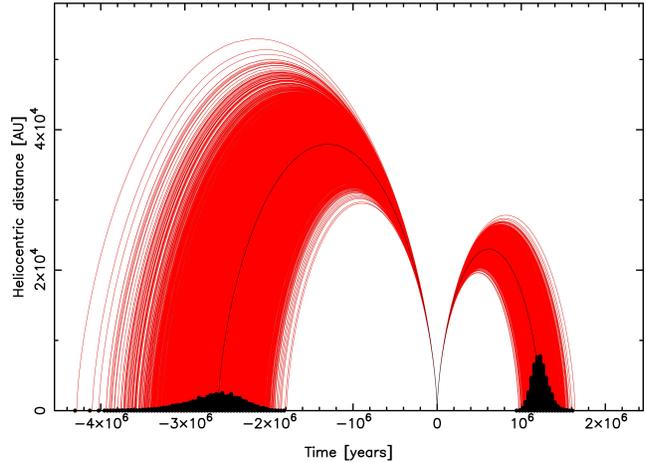}

\caption{\label{fig:fontanna_1913y1}Past and future motion of all 5001 VCs
for C/1913 Y1 in terms of their heliocentric distance. Solid, central
black line depicts the nominal orbit and black histograms show the
density of VCs previous/next perihelion passages in time. The time axis
is centered (t=0) at the observed perihelion passage of this comet
in 1914.}

\end{figure}

Previous perihelion passage of this comet is determined with the
highest accuracy of a group of comets discovered before 1970. Our
calculations show that C/1913~Y1 was previously inside inner Solar
System and lived it in an orbit of ${\rm q}_{\rm prev}<$3\,AU and
${\rm Q}_{\rm prev}<$43\,000\,AU (Table~\ref{tab:past_motion}) This
is the second comet in our sample which was originally considered by
Oort as dynamically new object. It seems worth to mention here, that
such a small previous perihelion distance was obtained as a result
of non-gravitational forces incorporated in this paper for the
\textbf{\emph{original}} and \textbf{\emph{future}} orbit
determination. Pure gravitational model gives the previous
perihelion value of the order of 150 AU for this comet.

The future motion of C/1913~Y1 is also determinable with very high
accuracy. In contrary to C/1885~X1 here all VCs are returning with
the distribution of the next perihelion and aphelion distances
only slightly departing from the Gaussian one. However, by fitting
to Gaussian one can derive the values of 0.97\,AU\,$\pm$0.02\,AU and
(23.1$\pm$1.1)$\times10^{3}$\,AU for next perihelion and aphelion
distance, respectively (compare with Table~\ref{tab:future_motion}).
The heliocentric motion of all VCs for one past and one future
orbital periods is shown in Fig.\ref{fig:fontanna_1913y1}.

If we demand, that the Oort cloud comet should move outside the sphere
of planetary perturbations (before it is observed for the first time)
C/1913 Y1 do not fulfill this condition either in the past or in future.

\subsubsection{Comets C/1892~Q1, C/1952~W1 and C/1959~Y1}
\label{sub:comets_discovered_before}

These three old comets have the past swarms of VCs widely spread.
Apart from a large number of escaping VCs, the significant part of
them have hyperbolic orbits. While in all cases we can call them
'dynamically new' with rather large degree of certainty, the exact
provenience of these comets is hard to be determined, we cannot
exclude even their interstellar origin. The probability of such a
case was extensively discussed by \citet{hughes:1991}.

\noindent C/1952 W1 is the only comet in our sample which have
formal value of $1/a_{\rm ori}<0$ (see Table~\ref{tab:past_motion}).
The future motion of C/1892~Q1 and C/1959~Y1 show evident hyperbolic
ejection while for C/1952~W1 64\% of VCs are escaping on hyperbolic
orbits but on the other hand some 30\% of VCs will return, sometimes
with very small perihelion distances (see
Table~\ref{tab:future_motion}).




\subsubsection{C/1986 P1 Wilson}\label{sub:C/1986-P1-Wilson}

This is very difficult comet to investigate because of its splitting
into two fragments. According to MWC08, the main component A of this
comet exhibits NG~acceleration in the time interval 1986\,August\,05
-- 1989\,April\,11. Therefore, we also determined the NG~effects
from the whole observational arc (see Table~\ref{tab:Obs-mat}) and
next starting from these nominal orbits (GR and NG) the past and
future orbital evolution have been analyzed. It turns out that the
fragment~B was a small piece of parent body and had ceased all
activity or had disintegrated after April 1989. To estimate the
possible uncertainty connected with this splitting we tried to find
the nominal orbits from two separate sets of data: before and after
splitting, but the exact date of this break-up is not clear. During
comet's observation on 1988\,February\,13 was discovered that
nucleus was fragmented. However, \citet{meech-etal:1995} argued that
the most preferable date for splitting was mid-October 1987 because
of the likely connection with a brightness of outburst in that time.
Thus, we decided to determine the past orbit of C/1986~P1 from the
time interval 1986\,August\,05 -- 1987\,July\,04 (next astrometric
observation was taken three months later on October 2) and the
future orbit from the time interval 1987\,October\,26 --
1989\,April\,11 (114 observations). We derived values of
$(53.14\pm4.14)\times10^{-6}$\,AU$^{-1}$ and
$(30.12\pm1.84)\times10^{-6}$\,AU$^{-1}$ for $1/{\rm a}_{{\rm
ori,NG}}$ and $1/{\rm a}_{{\rm ori,GR}}$, respectively, where NG
solution was determined with rms$=$1\farcs 30 and pure gravitational
solution -- with rms$=$1\farcs 35. In this approach we derived
37.6$_{-3.4}^{+4.1}\times10^{3}$\,AU and 1.90$_{-0.22}^{+0.38}$\,AU
for previous aphelion and perihelion passage, respectively, where by
adding the lower and upper errors to nominal value the deciles of
10\% and 90\% can be derived and directly compared with respective
values in Table~\ref{tab:past_motion}. By comparing both results of
previous perihelion swarms we stated that the comet C/1986~P1 were
deeply inside inner solar system with perihelion not larger than
4\,AU and aphelion less than $5\times10^{4}$\,AU.

\begin{figure}
\includegraphics[angle=270,width=1\columnwidth]{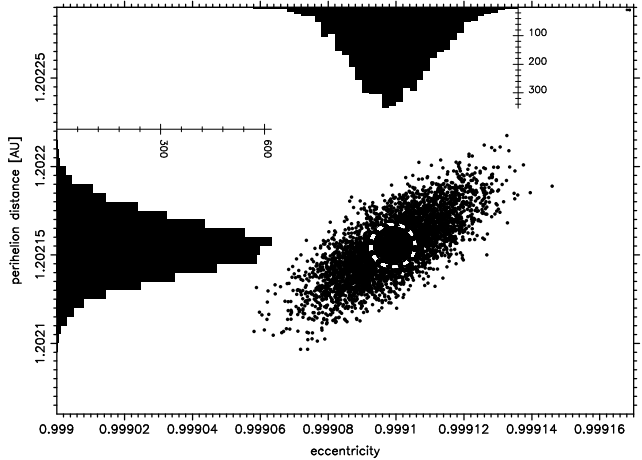}

\caption{\label{fig:1986p1fut}Future swarm for C/1986 P1. Note how small differences
appear between VCs during the next perihelion passage roughly 55\,000
year from now. The centre of the dashed-line circle marks the nominal
orbit. See text for additional comments.}

\end{figure}

For the data taken after the splitting the NG~effects were indeterminable.
The VCs swarm of future motion is the most compact in our sample,
what can be observed in Fig.\ref{fig:1986p1fut}. We derived the value
of $(749.48\pm9.83)\times10^{-6}$\,AU$^{-1}$ for $1/{\rm a}_{{\rm fut,GR}}$
By comparing this value with result given in Table~\ref{tab:future_motion}
we conclude that this comet will start subsequent perihelion passage
with small q (approximately 1.2\,AU) and with aphelion distance Q$\simeq$3\,000\,AU,
very similar to future aphelion distance of Comet C/1996~B2~Hyakutake.

\subsubsection{C/1989 Q1 Okazaki-Levy-Rudenko and C/1989~X1~Austin}
\label{sub:C/1989-Q1-Okazaki-Levy-Rudenko}

Both comets were observed in short time intervals (four and six months,
respectively) thus the past and future perihelion swarms of VCs are
relatively disperse what gives the formal chance of only
about 25\% that these comets are dynamically new. \\
 Comet C/1989~Q1 have a great majority (99\%) of returning VCs with
previous perihelion distance of the order of 1~AU but there is a
long tail in its distribution, consisting of 51 escaping VCs,
including 6 hyperbolas. Thus, the resulting $q$-deciles are
0.41\,AU--0.55\,AU--33.4\,AU (at 10\%, 50\%, and 90\%) and taking
limit $q{}_{{\rm previous}}=$15\,AU for dynamically new comets the
probability of 24.4\% was formally obtained that this comet is
dynamically new. \\
 Similarly, the $q$ deciles of 1.7\,AU--5.9\,AU--49.4\,AU (at
10\%, 50\%, and 90\%) give the chance of 25.9\% that comet C/1989~X1
is dynamically new.


\subsubsection{C/1990 K1 Levy}\label{sub:C/1990-K1-Levy}

This comet may serve as an excellent illustration, how the
incorporation of NG forces into the process of orbit determination
shrinks the semi-major axis and significantly shorten the time
interval of the past motion. When we use pure gravitational solution
for this comet (see the on-line extended table,
\citealt{dyb-krol:2009}) only 3644 VCs are classified as returning
while 1357 as escaping but elliptical. For the slowest returning VC
it takes almost 15$\times 10^6$\,years to reach the previous
perihelion. If we want all VCs to become returning, the 'escape
limit' should be increased up to 190\,000 AU what seems
unacceptable. But it is worth to mention that in that case the
slowest VC takes over 27 million years to pass the previous
perihelion.

>From Table~\ref{tab:past_motion} one can read, that when we use
NG~model, all VCs are returning with the standard 'escape border'.
Moreover, the slowest VC pass the previous perihelion point just
1.25$\times 10^6$\,years ago!

\subsubsection{C/1991 F2 Helin-Lawrence and C/1993 Q1 Mueller}
\label{sub:C/1991-F2-Helin-Lawrence}

These comets seem to be classical jumpers from the Oort Cloud with
largest $\Delta{\rm q}={\rm q}_{{\rm previous}}-{\rm q}_{{\rm ori}}$
among considered objects (larger than 300\,AU).

Since the astrometric set of data for C/1991~F2 consists of just 114
observations this comet represents a limit case for weighting. We
decided to show the past and future motion on the basis of weighted
observations (Tables~\ref{tab:past_motion} and
\ref{tab:future_motion}). It should be noted however, that in the
non-weighted data case the past swarm of VCs are almost all
returning (4968 of 5001 VCs) with q range:
22.3\,AU--71.9\,AU--340.8\,AU (three deciles at 10\%, 50\%, and
90\%) what gives 98.8\% of dynamically new VCs. Thus, the main
conclusion that this comet is dynamically new remains unchanged,
independently of data treating.

The past swarm of comet C/1993~Q1 is dominated by escaping VCs with
maximum of aphelion distance $\sim$150\,000\,AU (compare with
returning part of swarm given in Table~\ref{tab:past_motion}). Our
results suggest that this comet at previous perihelion was extremely
far from the Sun.

Both comets will be probably lost in the future -- they leave the
solar system on hyperbolic orbits.


\subsubsection{C/1996 E1 NEAT}\label{sub:C/1996-E1-NEAT}

Because only 6 VCs of this comet crossed the adopted escape limit in
the past motion we decided to increase this limit up to 140000 AU
when performing backward numerical integration. Then, all 5001 past
VCs were returning and 74\% of them had q$<$15~AU. After the observed
passage through the Solar System its nominal semi-major axis is
drastically reduced from 54\,000~AU down to 2600~AU and it will pass
the perihelion 1.3 AU from the Sun.

\subsubsection{Comets with large observed perihelion distance}
\label{sub:comets_with_largeq}

In our sample there are four comets with relatively large osculating
perihelion distances: C/1997~BA$_{6}$ Spacewatch, C/1997~J2
Meunier-Dupouy, C/1999~Y1 LINEAR and C/2000~SV$_{74}$ LINEAR. First
three might be grouped together to show different dynamical
histories of these comets in spite of their very similar original
perihelion distances (from 3.05\,AU to 3.44\,AU) and rather similar
original semi-major axes (within the range from 21\,000\,AU to
31\,500\,AU, Tab~\ref{tab:a_original}). One can see in
Tables~\ref{tab:past_motion}--\ref{tab:future_motion} that the
previous and future perihelion passages are well determined for all
of them. Thus, we can state that only the first one
(C/1997~BA$_{6}$~Spacewatch) having previous perihelion distance
around 20.0\,AU$\pm$3.3\,AU seems be dynamically new, whereas
remaining two comets were deep in the inner Solar System during the
past perihelion passage (Table~\ref{tab:past_motion}). This
may serve as an additional argument, that even having all parameters
of the original orbit one cannot judge what was the previous
aphelion or perihelion distance of a particular comet. While the
strength of the Galactic perturbations depends mainly on the
semi-major axis value, it can be significantly different, depending
on the orientation of the orbit with respect to the Galactic
equator. There exist another point, from
which C/1997 J2 Meunier-Dupouy is also an interesting comet. Its
previous perihelion distance (2.80\,AU) was slightly smaller than the original one (3.05\,AU),
what means that this comet was discovered during the {}``growing''
phase of its perihelion distance evolution under the influence of
the Galactic tides (two more comets in our sample,
C/1978~H1 and C/1990~K1 have also smaller previous
perihelion distance than the observed one). The statistical analysis
of the $q$ evolutional phase of long-period comets at their
discovery epoch was performed for example by
\citet{matese-lissauer:2004}. They concluded that most comets should
be observed during the {}``decreasing'' phase but noted several
opposite situations. It is worth to mention however, that their
analysis was based on pure gravitational cometary original orbits.

\noindent Future history of C/1997~J2 is more uncertain. It can be
seen from Table~\ref{tab:future_motion} that next VCs aphelion
distances are slightly larger than our assumed standard outer
border: 97\% of VCs are classified as escaping, but formally all of
them are elliptic, see Fig.\ref{fig:fut_1997j2syn}. By shifting the
escape border to 170\,000\,AU we can obtain all VCs returning and
approximately Gaussian distribution of aphelion distance
($Q=136.3$$\pm$8.4~thousand~AU) and non-Gaussian for perihelion
distance with values of 5.6\,AU, 7.9\,AU, 15.9\,AU for 1$^{{\rm
st}}$, 5$^{{\rm th}}$ and 9$^{{\rm th}}$ decile of $q$,
respectively. But so large aphelion distance means that this comet
will probably leave the Solar System. As it can be noted from
Table~\ref{tab:future_motion}, two remaining comets,
C/1997~BA$_{6}$~Spacewatch and C/1999~Y1~LINEAR, will have a very
similar future evolution. Both of them in the subsequent perihelion
passage will enter the inner Solar System in orbits with perihelion
distances between 3\,AU and 4\,AU and aphelion of
$\sim$5\,000\,AU--6\,000\,AU.

Fourth comet with large osculating perihelion distance, C/2000~SV$_{74}$~LINEAR,
have significantly shorter original semi-major axis of about 10\,800\,AU
and its previous perihelion passage was below 4\,AU. However, it
seems that C/2000~SV$_{74}$~LINEAR will leave the solar system
on a hyperbolic orbit.






\subsubsection{C/2001 Q4 NEAT and C/2002 T7 LINEAR}
\label{sub:well_determined_comets_1}

These comets were at great distance from the Sun when discovered
(comet C/2001~T7: over 11\,AU, C/2002\,T7: $\sim$7\,AU) and were
almost stellar in appearance. At maximum brightness both became easy
binocular objects (4\,mag. and 2.5--3\, mag, respectively) and as
consequence were observed long after perihelion passages. Such a
long observational arcs allowed to determine their orbits with
highest precision and we obtained very accurate results with
Gaussian distribution of VCs parameters both for past and future
motion.

Both comets have past swarms of VCs fully returning and compact,
however they differ dramatically in previous perihelion distance.
Comet C/2001~Q4 was at previous perihelion close to the Sun
(1.87\,AU$\pm$0.06\,AU), whereas C/~2002~T7 passed the previous
perihelion at a distance of 142\,AU$\pm$19\,AU from the Sun. Thus,
from this pair, only C/2002~T7 seems to be dynamically new.

While for these comets different form of the $g(r)$ functions might
be more appropriate, we are convinced\footnote{We have tested
various shapes of others g(r)-like functions (Eq.\ref{g_r}) for both
comets and differences in comet's osculating orbital elements among
all considered NG~models were about one order of magnitude smaller
than obtained between gravitational and NG models} that the main
conclusions remains the same.

Both comets are leaving the Solar System on hyperbolic orbits.


\section{Conclusions}\label{sec:Conclusions}

The aim of our research is to refine widely quoted opinions about
the apparent source of the long period comets and a popular habit to
call some of them \textbf{\emph{dynamically new}} by looking on the
original inverse semi-major axis value only.

In this paper, the first one of a planned series, we touch the most
important group of comets, 26 long-period comets with $1/{\rm
a}_{\rm ori}<10^{-4}$\,AU$^{-1}$ for which we can obtain very
precise orbits including NG forces in the dynamical model. The
importance of this sample comes from the fact that for almost all of
these comets we obtained significantly smaller semi-major axes due
to NG force incorporation in the process of orbit determination, but
still all of them constitute the NG~Oort spike, i.e have $1/{\rm
a}_{{\rm ori,NG}}<10^{-4}$\,AU$^{-1}$.

Having such a precise orbits we applied ingenious Sitarski's
\citeyearpar{sitarski:1998} method of creating swarms of VCs, all
compatible with observations. This allowed us first to obtain
\textbf{\emph{original}} and \textbf{\emph{future}} orbits
parameters with their uncertainties and then to propagate all
individual VCs one orbital period to the past and to the future, by
means of strict numerical integration of the equation of motion with
the Galactic tide action included.

Note, that our neglecting of stellar perturbations is based on well
documented arguments \citep{dyb-hab3:2006}, but the incorporation of
NG~forces into the orbit determination makes these arguments even
stronger. All but one of comets in our sample (and many others,
see \citealp{krolikowska:2006a}) appear to have significantly smaller
semi-major axes what implies that their orbital period becomes
drastically shorter. This makes a strong stellar perturbation by an
undiscovered massive and slow moving star extremely improbable.

Our results show, that two widely quoted opinion about the
long-period comets are not necessary the {}``canonical truth''.
First, all 26 comets analyzed in this paper have gravitational
inverse of the semi-major axis very small ($1/{\rm a}_{\rm
ori,GR}<6\times 10^{-5}$\,AU$^{-1}$) what makes almost all authors
to call them \textbf{\emph{dynamically new}}. But from
Table~\ref{tab:past_motion} one can clearly recognize, that no more
than 5 or 6 of them deserve for such a name. The rest visited our
Planetary System during the previous perihelion passage, possibly
experiencing strong planetary perturbations and nobody knows what
process or mechanism direct them at that time (or later) to the
observable orbit at which we discovered them.

Second, we showed, that about 30\% of our sample of long-period
comets will remain observable in the next perihelion passage.
Assuming that this is not a strange statistical fluctuation and
keeping in mind that the only recognized force which makes their
orbits to evolve is the tidal action of the Galactic matter, the
widely quoted necessary condition for a comet to be made observable
by Galactic perturbation is too strong. It is typical to demand that
such a comet should decrease its perihelion distance from above 15
AU down to the observational value in one orbital period, otherwise
they cannot cross the so called Jupiter-Saturn barrier, what now
seems to be a {}``not necessary'' condition:

\textbf{\emph{We have shown that a significant percentage of long-period
comets can visit the zone of visibility during at least two or three
consecutive perihelion passages.}}

In this context we would like to suggest some correction to the
widely quoted and very useful comet classification scheme proposed
by \citet{levison:1996}. In his taxonomy the nearly isotropic comets
(NICs) are divided into two subpopulations: new NICs and returning
NICs on a basis of their original semi-major axis. This division
should probably be replaced by a new one, based on the past
dynamical history of a particular comet, especially if we want to
keep naming NICs 'new' or 'returning' from the dynamical point of
view.

There exist also a difficult problem of linking the 'dynamical age'
of long-period comets with their physical properties, including the
chemical composition \citep{cro:2007,cro:2009,Biver:2002}. We tried
to look for such a linkage but it seems that physical and dynamical
characteristics might be quite different. For example the spectra of
C/2001~Q4 NEAT and C/2002~T7 LINEAR were intensively investigated as
they were bright and well observed comets. The observations of large
molecules in both comets show \citep{Remijan:2006} that they are
chemically very similar while our 'dynamical ages' for these comets
are completely different. However, \citet{Remijan:2006} also stated
that when comparing the molecular production ratios with respect to
water it appears that C/2002~T7 is more similar to Hale-Bopp
physical class of comets (comets rich in HCN: HCN/H$_{2}$O$>0.2$\%),
while C/2001~Q4 is more similar to Hyakutake class  (comets with
HCN/H$_{2}$O abundance ratio of $\sim$0.1\%) where two more
investigated here comets (C/1989 X1 and C/1990~K1) are also
included. On the other hand, C/1995~O1 Hale-Bopp and C/1996~B2
Hyakutake seems to be dynamically more similar than comets Hale-Bopp
and C/2002~T7.

>From the point of view of the apparent source of the long period
comets, the main result of our paper is the list of previous
aphelion and perihelion distances together with numbers of escaping
VCs, presented in Table~\ref{tab:past_motion}. But, comparing the past
and future (Table~\ref{tab:future_motion}) dynamics of
26~comets bring us to an interesting additional
conclusion.

Treating these results as a probe of the past and future flux of
observable long-period comets one can conclude that there exist
surprising discrepancy between past and future cometary orbit
evolution. Some 33\% of comets arriving at the outer border of the
Planetary System (250\,AU from the Sun) are dynamically new, thus
66\% have visited our Planetary System one orbital period before. On
the other hand, 65\% of comets leaving the zone of visibility will
be lost in interstellar space and only 35\% will return. But, in the
{}``next turn'', they should constitute 66\% of observable
long-period comets, what might suggest that the flux of long-period
comets should constantly decrease...

And the last remark, of more historical than scientific value. As
far as we are able to recognize now \citep{dyb-Web-Oort:2009}
only two comets from our sample studied in this paper were used by
\citet{oort:1950} in construction of his Table 1, namely C/1885 X1
Fabry and C/1913 Y1 Delavan. According to our results both these
comets have visited the Planetary System during their previous
perihelion and passed rather close to the Sun ( 1-2 AU)
approximately 2.5$\times 10^6$\,years ago.

As it concerns our future plans in this research, we have calculated
over 100 precise original and future orbits, waiting for the
detailed analysis of their past and future motion. In the next step
we probably concentrate on {}``statistics improving'' by significant
increasing of number of analyzed comets, with a special attention on
large perihelion distance comets, where even without detectable
NG~effects, orbits are very precise.

\section{Appendixes}

\subsection{Chauvenet's and Bessel criterions}\label{app:Cha}

Both criteria differ in the upper limit of the accepted residuals,
$\xi$, e.g. observed minus computed values of right ascension, $\Delta\alpha\cdot\cos\delta$,
and declination, $\Delta\delta$. According to the Chauvenet's criterion
\citep{cha:1908} from the set of $N$ residuals, $\xi$, we should
discard all values of $\xi$ for which

\[
\mid\xi\mid>\sigma\cdot K_{1/2}(N)\]
 where $\sigma$ is a dispersion of $\xi$:

\[
\sigma=\sqrt{\left(\sum_{k}\xi_{k}2\right)/N}\]
 and $K_{1/2}(N)$ is the unknown upper limit of the integral of the
probability distribution, $\phi(\xi)$:

\[
\int_{0}^{K_{1/2}}\phi(x){\rm d}x=1-\frac{1}{2N}\,,\]
 where $x=\xi/\sigma$.

According to this criterion the data point is rejected if the probability
of obtaining the particular deviation of residuals from the mean value
is less than $1/(2N)$. To determine this probability the normal distribution
of $\xi$ is assumed.

\noindent The Bessel criterion (more restrictive than the Chauvenet's
criterion) rejects from the set of $N$ residuals all the values of
$\xi$ for which

\[
\mid\xi\mid>\sigma\cdot K_{1}(N)\,,\]
 where $K_{1}(N)$ is defined by:

\[
\int_{0}^{K_{1}}\phi(\xi){\rm d}\xi=1-\frac{1}{N}.\]

\section*{Acknowledgments}

We would like to thank many participants of numerous fruitful
discussions during preparation of this paper. We would like to
express our special thanks to Grzegorz Sitarski, Hans Rickman and
Giovanni Valsecchi. Detailed and important suggestions of the
anonymous referee allow us to improve the quality of this paper. The
research described here was partially supported by Polish Ministry
of Science and Higher Education funds (years 2008-2010, grant no. N
N203 392734) This manuscript was partially prepared with \LyX{}, the
open source front-end to the \TeX{} system.

\bibliographystyle{mn2e}

\bibliography{moja13}

\label{lastpage}

\end{document}